\documentclass[12pt]{article}
\pdfoutput=1 
\usepackage[dvipdfmx]{graphicx}
\usepackage{amsfonts,amsmath,amssymb,epsf}
\usepackage{amsmath,braket}
\usepackage{graphicx,color}
\newcommand{\al}{\alpha'}
\newcommand{\de}{\partial}
\newcommand{\be}{\begin{equation}}
\newcommand{\ba}{\begin{eqnarray}}
\newcommand{\ea}{\end{eqnarray}}
\newcommand{\ee}{\end{equation}}

\newcommand{\s}{\sqrt}
\newcommand{\vp}{\varphi}

\newcommand{\ti}{\tilde}

\newcommand{\ddd}{\cdot\cdot\cdot}
\newcommand{\no}{\nonumber \\}
\newcommand{\la}{\langle}
\newcommand{\lb}{\rangle}
\newcommand{\bea}{\begin{eqnarray}}
\newcommand{\eea}{\end{eqnarray}}
\newcommand{\bes}{\begin{equation*}}
\newcommand{\beas}{\begin{eqnarray*}}
\newcommand{\eeas}{\end{eqnarray*}}
\newcommand{\bas}{\begin{array*}}
\newcommand{\eas}{\end{array*}}
\newcommand{\ees}{\end{equation*}}

\newcommand{\ep}{\epsilon}

\usepackage{cite}

\begin{document}

\begin{titlepage}
\thispagestyle{empty}

\begin{flushright}
YITP-19-116
\\
IPMU19-0173
\\
\end{flushright}

\bigskip

\begin{center}
\noindent{{\large \textbf{Gravity Edges Modes and Hayward Term}}}\\
\vspace{2cm}

Tadashi Takayanagi$^{a,b}$ and Kotaro Tamaoka$^{a}$
\vspace{1cm}\\

{\it $^a$Center for Gravitational Physics,\\
Yukawa Institute for Theoretical Physics,
Kyoto University, \\
Kitashirakawa Oiwakecho, Sakyo-ku, Kyoto 606-8502, Japan}\\

{\it $^{b}$Kavli Institute for the Physics and Mathematics
 of the Universe (WPI),\\
University of Tokyo, Kashiwa, Chiba 277-8582, Japan}

\end{center}

\begin{abstract}
We argue that corner contributions in gravity action (Hayward term) capture the essence of gravity edge modes, which 
lead to gravitational area entropies, such as the black hole entropy and holographic entanglement entropy. 
We explain how the Hayward term and the corresponding edge modes in gravity are explained by holography
from two different viewpoints. One is an extension of AdS/CFT to general spacetimes and the other is the AdS/BCFT formulation.
In the final part, we explore how gravity edge modes and its entropy show up in string theory by considering open strings 
stuck to a Rindler horizon.
\end{abstract}

\end{titlepage}

\newpage

\section{Introduction}

Gravitation theory has the remarkable feature that a spacetime can have a non-zero entropy even at the classical theory level.
There is no such classical entropy in theories of matter fields, such as scalar and fermion fields. The most well-known manifestation 
of this special property of gravity is the black hole entropy \cite{Bekenstein:1973ur,Hawking:1974sw}, given by the celebrated formula:
\ba
S_{BH}=\frac{A(\Gamma_{BH})}{4G_N},  \label{BHF}
\ea
where $A(\Gamma_{BH})$ is the area of the black hole horizon $\Gamma_{BH}$ and $G_N$ is the Newton constant.

Though this fact that the gravitational entropy is proportional to the area instead of the volume bothered researchers for a long time,
it was realized that this is actually the essence of gravity and even provides basic principle, i.e. holographic principle 
\cite{tHooft:1993dmi,Susskind:1994vu}. Moreover, string theory, as the best candidate of quantum gravity, offered a microscopic 
explanation of black hole entropy as pioneered in \cite{Strominger:1996sh}. These developments finally lead to the concrete formulation of 
holography, namely the AdS/CFT (anti de-Sitter/conformal field theory) correspondence \cite{Ma,Gubser:1998bc,Witten:1998qj}. In this approach, the black hole entropy gets simply equal to thermodynamical entropy in the dual CFT, which lives in a one dimension lower spacetime.

In fact, gravitational entropy arises even without black holes.  In AdS/CFT, entanglement entropy in CFTs for a subsystem $A$ can be computed as the area of minimal surface\cite{RT,HRT}
\ba
S_A=\frac{A(\Gamma_A)}{4G_N},  \label{RTF}
\ea
where $\Gamma_A$ is the minimal area surface in the AdS which ends on $\de A$ at the AdS boundary.  We can view this holographic entanglement entropy formula (\ref{RTF}) as a generalization of the Bekenstein-Hawking formula (\ref{BHF}), where the former is reduced to the latter when $A$ is the total system in the presence of black holes.

Moreover, a more generalized area entropy formula was conjectured in \cite{Bianchi:2012ev}. It was argued that an area of any space-like surface $\Gamma$ in a gravity background is equal to entanglement entropy in gravity which measures quantum entanglement between the region $\Gamma$ and the outside on a suitable time slice in gravity. We may think this as a generalized conjecture of (\ref{BHF}) and (\ref{RTF}). 

In this way, we have now had abundant understandings of gravitational entropy from the viewpoint of holography or AdS/CFT correspondence,
where the classical gravity entropy is explained by the deconfined degrees of freedom in CFTs.  However, this is an indirect argument and
we are still not be able to answer the origin of classical gravitational entropy in terms of gravity theory itself.  For example, we still need to ask how we can identify such an entropy in the Hilbert space of quantum gravity. Even though the bulk theory does not seem to include the degrees of freedom which explain the $O(1/G_N)=O(N^2)$ classical entropy as the number of fields in supergravity is $O(1)$, the presence of gravitational entropies (\ref{BHF}) and (\ref{RTF}) argues that there is actually $O(1/G_N)=O(N^2)$ entropy localized on a surface $\Gamma$. 
In this sense, this situation in gravity may look slightly analogous to topological ordered systems, where gapless degrees of freedom appear on boundaries, and 
therefore the classical contribution to the gravitational entropy is often called ``gravity edge modes''.  Refer to \cite{Donnelly:2016auv,Speranza:2017gxd,Geiller:2017xad,Camps:2018wjf, Harlow:2015lma,Harlow:2018tqv,Casini:2019kex,Benedetti:2019uej,Lin:2018xkj,Jafferis:2019wkd, Harlow:2016vwg, Lin:2017uzr, Akers:2018fow, Dong:2018seb, Dong:2019piw} for recent discussions. 

The purpose of this paper is to explore more on gravity edge modes from the viewpoints of both the AdS/CFT and string theory. 
We will spend a more than half part of the present paper to point out that the corner contribution to the gravity action \cite{Hayward:1993my} (see also \cite{Hayward:1992ix,Brill:1994mb} for more aspects and \cite{Lehner:2016vdi, Takayanagi:2018pml,Sato:2019kik,Braccia:2019xxi} for applications to holographic complexity \cite{Brown:2015bva}), which we call Hayward term, captures the essence of the gravity edge modes. This term arises the spacetime includes a non-smooth boundary 
which we call a wedge. In other words, the Hayward term is a codimension two analogue of  the Gibbon-Hawking term \cite{Gibbons:1977mu}.
We will divide the original gravitational spacetime into multiple regions and 
consider pasting them together where the non-additive nature of gravity action \cite{Brill:1994mb} plays a crucial role. There are two 
different interpretations in terms of AdS/CFT. One is to impose the Dirichlet boundary conditions on all boundaries of the divided spacetimes.
In this approach, the gravitational entropy of the edge modes is explained by the CFT dynamics on the extra boundaries which emerge by cutting the original spacetime.  We can equally interpreted the entropy as that for the purification \cite{UT,Nguyen:2017yqw}. The other one is to combine the Neumann boundary condition with Dirichlet one so that we can view each divided spacetime as a gravity dual of a boundary conformal field theory (BCFT) in the formulation of AdS/BCFT \cite{AdSBCFT} (see also earlier works \cite{Karch:2000gx}). In this interpretation, we can identify the gravitational entropy with the number of boundary conditions in BCFT.  In addition, we will show the Hayward term is necessary to reproduce the correct conformal anomaly in a class of setups of AdS/BCFT. Interestingly, we will show that each of these two approaches has an analogy in edge contributions in gauge theories. All these results strongly suggest that the essence of gravity edge modes is condensed into the Hayward term.

In the final part of this paper, we will explore a string theory origin of gravity edge modes. In the pioneering work \cite{Susskind:1994sm}, it was proposed that the classical gravity entropy should be explained by viewing the sphere partition function of closed string world-sheet where the two poles are stuck to Rindler the horizon as an open string partition function. We will present an explicit formulation to realize this idea. Finally we will show that the open string computation leads to the expected gravitational entropy up to an undetermined $O(1)$ constant. This strongly suggests string theory includes the degrees of freedom of gravity edge modes properly.

This paper is organized as follows. In section two, we will give a brief review of Hayward term and its derivation. After that we will 
derive the canonical formulation of the gravity edge modes. In section three, we first present the replica method computation of 
gravitational entropy by employing the Hayward term. Then we summarize the basic rules how the Hayward term appears in gravitational partition functions. In section four, we show how the Hayward term and gravitational entropy are explained in the context of holography.
In particular we provides two independent arguments based on a generalized holography and on the AdS/BCFT formulation. 
In section five, we present analogies of these two arguments in terms of entanglement entropy in gauge theories.    
In section six, we explore how gravity edge modes and its entropy show up in string theory by considering open strings stuck to a Rindler horizon. In section seven, we summarize our conclusions and discuss future problems.

\section{Gravity Action in the Presence of Wedges}

\begin{figure}
  \centering
  \includegraphics[width=6cm]{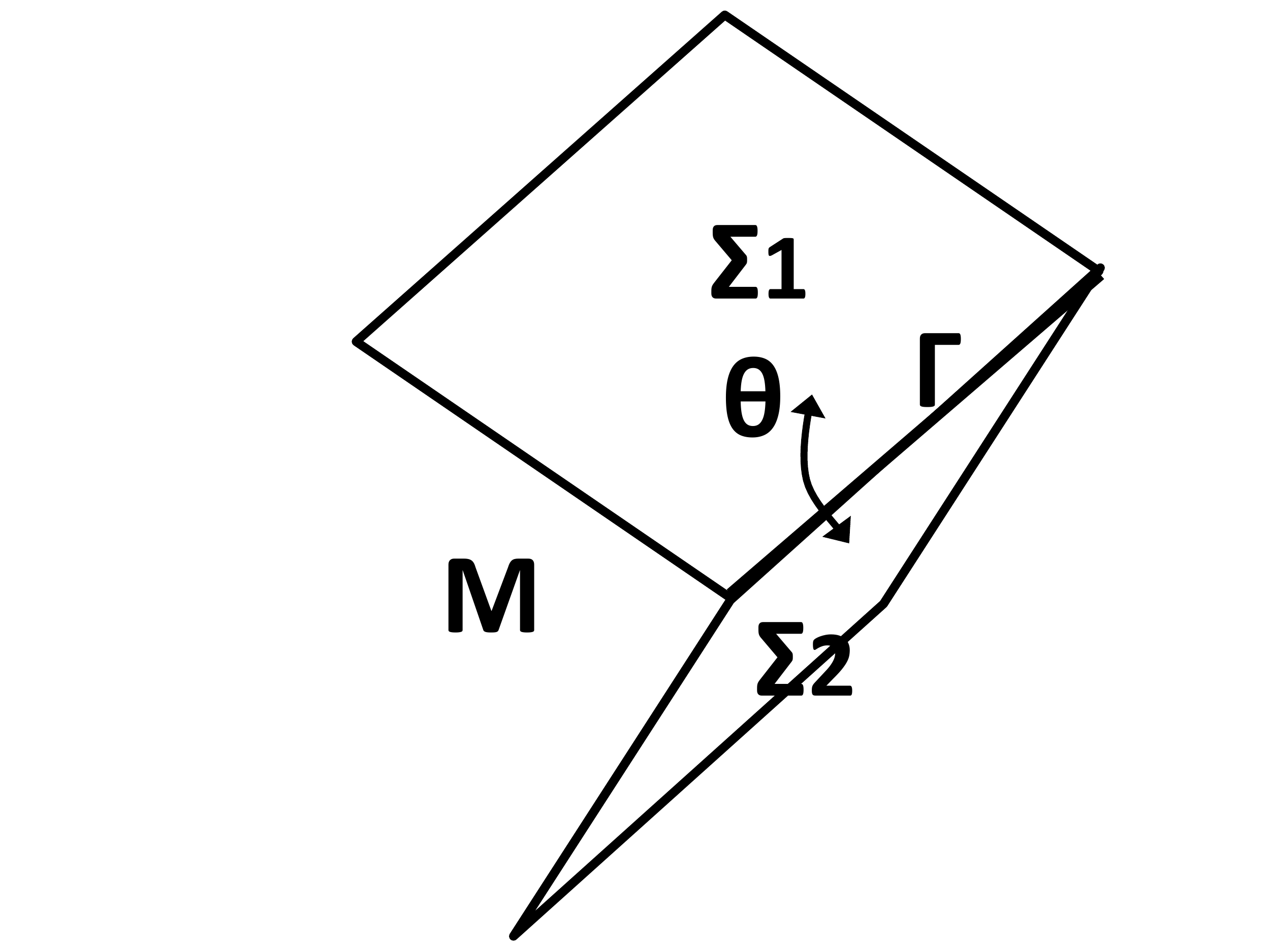}
  \caption{A sketch of Euclidean space $M$ with a wedge.  The wedge is situated along the co-dimension two surface $\Gamma$ with the angle $\theta$.}
\label{hay}
\end{figure}

Consider a gravity action on an Euclidean manifold $M$ in the the presence of a wedge as in Fig.\ref{hay}.
The wedge is bounded by the two surfaces $\Sigma_1$ and $\Sigma_2$  such that $\de M=\Sigma_1\cup \Sigma_2$, whose intersection (or corner) 
is called $\Gamma$. We write the (induced) metric on $M, \Sigma_{1,2}$ and $\Gamma$ by  $g, h$ and $\gamma$, respectively.

The full gravity action is expressed as
\ba
I_M&=&-\frac{1}{16\pi G_N}\int_M \s{g}(R-2\Lambda)-\frac{1}{8\pi G_N}\int_{\Sigma_1} \s{h}K
-\frac{1}{8\pi G_N}\int_{\Sigma_2} \s{h}K \no
&&+\frac{1}{8\pi G_N}\int_{\Gamma}(\theta-\pi)\s{\gamma},  \label{gac}
\ea
where $K$ is the trace of the extrinsic curvature; 
$\theta$ is the angle between the two surfaces $\Sigma_1$ and $\Sigma_2$. The final term 
\ba
I_{H}=\frac{1}{8\pi G_N}\int_{\Gamma}(\theta-\pi)\s{\gamma},  \label{hayward}
\ea
is the Hayward term \cite{Hayward:1993my} (refer also to \cite{Hayward:1992ix,Brill:1994mb}) which plays an important role in this article.

\subsection{A Sketch of Derivation of Hayward Term}

When we take the variation of the Euclidean gravity action (\ref{gac}) without the Hayward term, we obtain
\ba
\delta (I_M-I_H)&=&-\frac{1}{16\pi G_N}\int_M(R_{\mu\nu}-\frac{1}{2}Rg_{\mu\nu}+\Lambda g_{\mu\nu})
\delta g^{\mu\nu} -\frac{1}{8\pi G_N}\int_{\Sigma_1\cup \Sigma_2} \s{h}(K_{ab}-h_{ab}K)\delta h^{ab}\no
&&\ \ \  -\frac{1}{8\pi G_N}\int_{\Gamma}\s{\gamma}\delta\theta.
\ea
Therefore, after we impose the Einstein equation 
\ba
R_{\mu\nu}-\frac{1}{2}Rg_{\mu\nu}+\Lambda g_{\mu\nu}=0,
\ea
and the boundary counterpart 
\ba
K_{ab}-h_{ab}K=0,  \label{baein}
\ea
we need to set the $\Gamma$-boundary term to zero:
\ba
\s{\gamma}\delta\theta=0.  \label{constbh}
\ea

Consider the case where we impose the Dirichlet boundary condition  on $\Sigma_1$ and 
$\Sigma_2$ i.e. we fix the metric $h_{ab}(x)$ on $\Sigma_1\cup \Sigma_2$.  Then the 
boundary equation of motion (\ref{baein}) determines the angle $\theta$ dynamically.  Moreover, the Dirichlet boundary
condition fixes the induced metric $\gamma$. 
Therefore, it is not desirable to set $\theta$=fixed from the beginning. In this reason, we 
would like to avoid the boundary constraint (\ref{constbh}).

This motivates us to add the Hayward term (\ref{hayward}) and then the variation becomes
\ba
\delta I_M&=&-\frac{1}{16\pi G_N}\int_M(R_{\mu\nu}-\frac{1}{2}Rg_{\mu\nu}+\Lambda g_{\mu\nu})
\delta g^{\mu\nu} -\frac{1}{8\pi G_N}\int_{\Sigma_1\cup \Sigma_2} \s{h}(K_{ab}-h_{ab}K)\delta h^{ab}\no
&&\ \ \  +\frac{1}{8\pi G_N}\int_{\Gamma}(\theta-\pi)\delta\s{\gamma}.
\ea
This is sensible as the Dirichlet boundary condition on $\Sigma_1\cup \Sigma_2$ determines the value of 
$\s{\gamma}$ on $\Gamma$ at the same time and therefore leads to $\delta \s{\gamma}=0$.
The constant $\pi$ in $(\theta-\pi)$ is determined from the fact that there should be no Hayward term when the boundary is smooth at $\Gamma$ i.e. $\theta=\pi$.

\subsection{Hayward Term as Edge Modes}
Next let us study the Hayward term in a canonical formalism, which will reveal its nature as edge modes. 
Let us start from a $d+1$ dimensional spacetime $M$ which has no boundary (and wedge) for simplicity. Note that here we will discuss a Lorentzian spacetime. Let us use the canonical formalism for the gravity action to make dynamical variables in the wedge clear \cite{Hayward:1992ix}.
To do this, we use the standard ADM formalism. We first foliate the spacetime by the fixed $t$ timeslice as follows,
\ba
ds^2=g_{\mu\nu}dx^\mu dx^\nu=(-N^2+h_{ij}N^iN^j)dt^2+2h_{ij}N^jdx^idt+h_{ij}dx^idx^j,
\ea
where $h_{ij}$ is an induced metric of the time slice, $N$ is the lapse and $N^i$ is the shift function. 

\begin{figure}
  \centering
  \includegraphics[width=6cm]{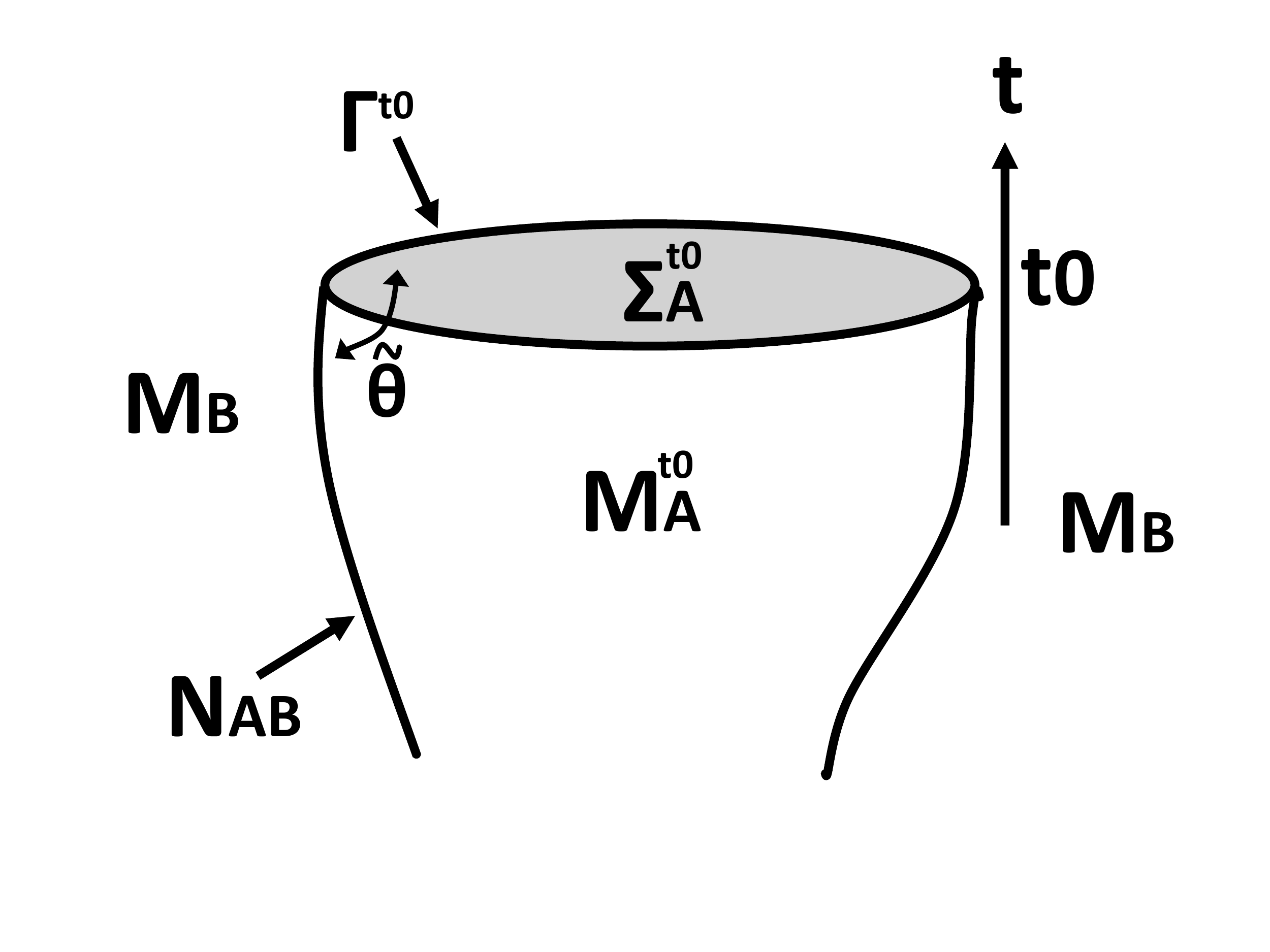}
  \caption{A sketch of Lorentzian gravity setup for the canonical formalism.}
\label{HayE}
\end{figure}

Next, we discuss the canonical formalism in a given subsystem. Suppose we split the spacetime $M$ into two pieces, say $M_A$ and $M_B$ along 
a time-like boundary $N_{AB}=\de M_A=\de M_B$. At a specific time $t_0$,  a time slice $\Sigma^{t_0}_A$ 
intersects with the boundary $N_{AB}$ along a $d-1$ dimensional surface $\Gamma^{t_0}$.
We consider the subspacetime $M^{t_0}_A$ defined by the lower half region of $M_A$ defined by the restriction $t\leq t_0$.
Refer to Fig.\ref{HayE} for this setup.

If we take a variation of the Lorentzian gravity action for $M^{t_0}_A$ including the Hayward term, we obtain
\ba\label{variation}
\delta I_{M^{t_0}_A} =\frac{1}{16\pi G_N}
\int_{M^{t_0}_A}\s{-g} (R_{\mu\nu}-\frac{1}{2}Rg_{\mu\nu}+\Lambda g_{\mu\nu})\delta g^{\mu\nu} +\int^{t_0}_{-\infty} dt \frac{d}{dt}\Theta_A(h_{ij},\gamma),
\ea
where
\be
\Theta_A(h_{ij},\gamma) = \int_{\Sigma^{t_0}_A} \pi^{ij}\delta h_{ij}  +\int_{\Gamma^{t_0}} \ti{\theta} \,\frac{\delta\sqrt{\gamma}}{8\pi G_N},  \label{cornert}
\ee
is so-called the symplectic potential which makes the degrees of freedom in the phase space manifest. Here $\pi^{ij}$ is a canonical momentum conjugate to the $h_{ij}$. 
We neglected boundary contributions other than the wedge term just for simplicity. The degrees of freedom described by $\ti{\theta}$ is the boost angle at the corner 
$\Gamma^{t_0}$, namely the boost angle in Lorentzian spacetime between $\Sigma^{t_0}_A$ and $N_{AB}$ along $\Gamma^{t_0}$. In the Wick rotation into the 
previous Euclidean signature, we can identify $\ti{\theta}=i\theta$.  Note that at the semiclassical level, the Hartle-Hawking wave function $\Psi_{HH}$ on the 
region $\Sigma^{t_0}_A$ for the spacetime $M_A$ is given by the above action as $\Psi_{HH}=e^{iI_{M^{t_0}_A}}$ 

The presence of the corner term (\ref{cornert}) in the canonical formulation implies that we have dynamical degrees of freedom on the wedge.  
The commutation relation for the area density $\sqrt{\gamma}$ and the boost angle $\tilde{\theta}$ on the codimension two surface $\Gamma$ is given by
\ba
[\,\tilde{\theta}(x), \sqrt{\gamma}(y)\,]=i8\pi G_N\delta(x-y). \label{eq:ccr}
\ea

To make the variational problem in \eqref{variation} well-defined, we need to impose extra boundary conditions on the wedge. Then, one obtains a well-defined classical system. In principle, one can quantize it for each boundary condition\cite{Donnelly:2016auv} (see also \cite{Speranza:2017gxd,Geiller:2017xad,Camps:2018wjf,Harlow:2018tqv} as examples for gravitational subsystems). This choice corresponds to one of the superselection sectors in the light of the operator algebra of a subregion\cite{Casini:2013rba}. 
Since we should fix the metric on the wedge, it is similar to the ``magnetic center'' in the algebraic definition of subsystems in gauge theories.
Note that for the derivation of the Hayward term, we chose explicitly the decomposition of original metric. It means that we partially fixed the gauge degrees of freedom normal to co-dimension 2 surface. We can also regard our edge modes' symplectic potential from Hayward term as a specific gauge fixing of discussion in \cite{Donnelly:2016auv}. In there, we have $SL(2,\mathbb{R})$ symmetry which is now spontaneously broken to the canonical commutation relation \eqref{eq:ccr}.

\section{Gravitational Entropy from Hayward Term}

Consider the replica calculation of entanglement entropy in gravity using the gravity action (\ref{gac}). 
Usually this is done by introducing a deficit angle along a codimension two surface, which is a horizon in the black hole case \cite{Fursaev:1995ef}
and which is a minimal surface in the holographic entanglement entropy \cite{Fursaev:2006ih}. Below we would like to present an equivalent computation 
in terms of wedges instead of deficit angles. This calculation will reveal the basic connection between the Hayward term and gravitational entropy.
Our analysis is general in that the codimension two surface $\Gamma$ can be chosen arbitrarily in the whole Euclidean space, which fits nicely with the conjecture 
\cite{Bianchi:2012ev}. In the context of holographic entanglement 
entropy in AdS/CFT we have in mind the area fixed state calculation \cite{Fursaev:2006ih} instead of the CFT vacuum \cite{Lewkowycz:2013nqa} in the sense of 
\cite{Dong:2018seb,Akers:2018fow}. After we present the replica calculation of gravitational entropy, we will explain general rules of semiclassical computations of the gravity 
partition functions in the presence of Hayward term.

\subsection{Replica Calculation using Hayward Term}

\begin{figure}
  \centering
  \includegraphics[width=6cm]{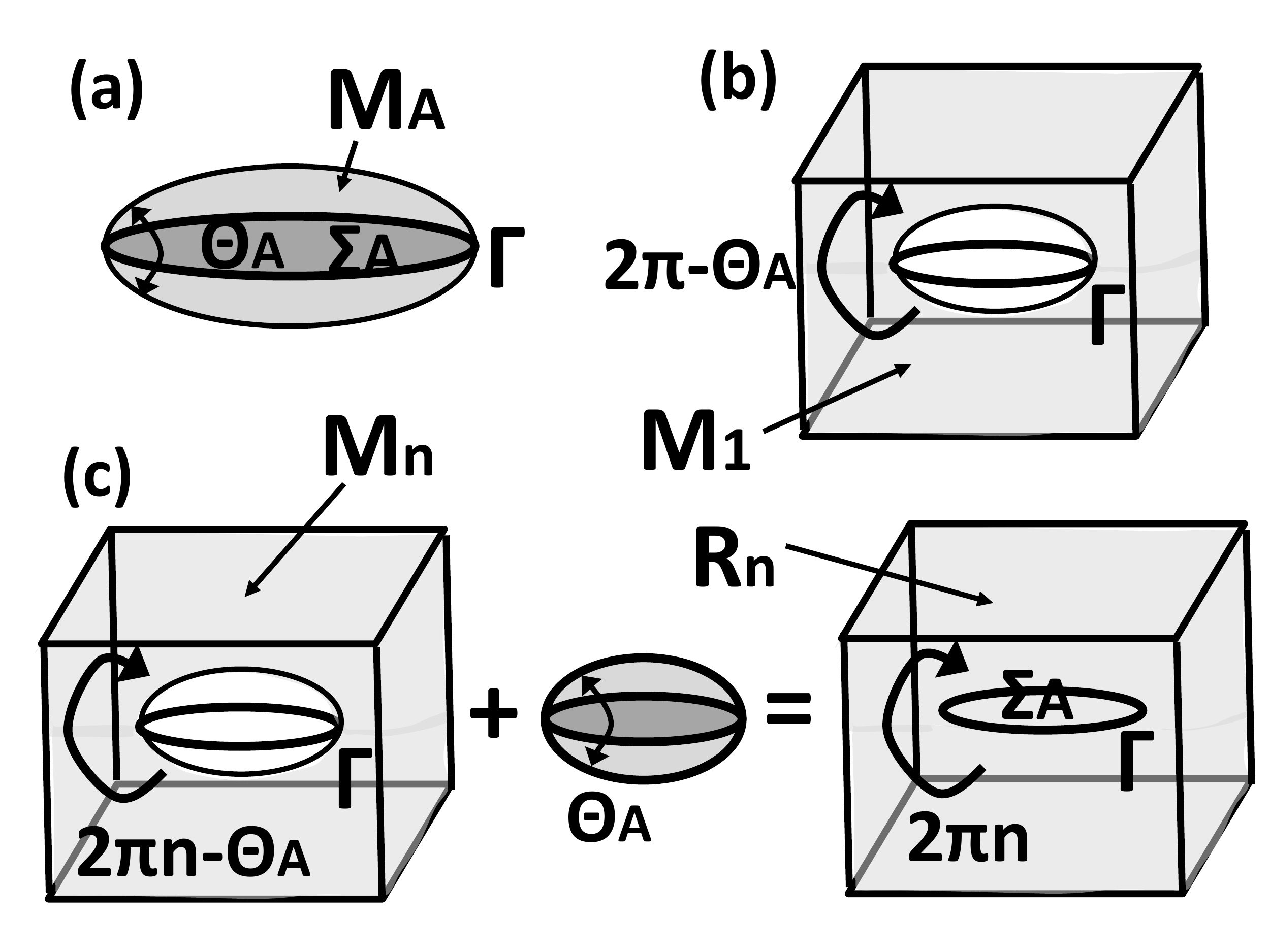}
  \caption{Sketches of Replica Calculations of Gravitational Entropy.}
\label{replicafig}
\end{figure}

We assume the gravity is originally defined in the flat space $R^{d+1}$ for simplicity. 
In the classical gravity limit, we can semiclassically evaluate the gravity partition function $Z_M$ on $M$ from the on-shell gravity action $I_M$ as 
\be
Z_M=e^{-I_M}.  \label{graac}
\ee 
We consider the entanglement entropy $S_{\Sigma_A}$, where  a time slice is divided into  two subregions $\Sigma_A$ and its compliment. This division is specified by the surface $\Gamma$ such that $\de \Sigma_A=\Gamma$. The calculation which we will explain below is sketched in Fig.\ref{replicafig}.

First we consider a small space described by (a) in  Fig.\ref{replicafig}. This is a region, called $M_A$ with a wedge along $\Gamma$ with the angle $\theta_A$. The gravity action on $M_A$ is evaluated as 
\ba
I_{M_A}=-\frac{1}{16\pi G_N}\int_{M_A} \s{g}(R-2\Lambda)-\frac{1}{8\pi G_N}\int_{\de M_A} \s{h}K
+\frac{1}{8\pi G_N}\int_{\Gamma}(\theta_A-\pi)\s{\gamma}.  \label{maac}
\ea

On the other hand we introduce a space $M_1$ i.e. (b) in  Fig.\ref{replicafig}.  This is defined by eliminating 
the space $M_A$ from the original space manifold $R^{d+1}$.
The gravity action on $M_1$ reads
\ba
I_{M_1}=-\frac{1}{16\pi G_N}\int_{M_1} \s{g}(R-2\Lambda)-\frac{1}{8\pi G_N}\int_{\de M_1} \s{h}K
+\frac{1}{8\pi G_N}\int_{\Gamma}(\pi-\theta_A)\s{\gamma}.  \label{mbac}
\ea

Therefore it is clear that the sum of the two (by cancellations of extrinsic curvatures in both sides) 
\ba
I_{MA}+I_{M1}=-\frac{1}{16\pi G_N}\int_{R^{d+1}} \s{g}(R-2\Lambda),
\ea
coincides with the gravity action on the original space $R^{d+1}$. From the partition function viewpoint 
we find the obvious relation
\ba
Z_{R^{d+1}}=Z_{M_A}\cdot Z_{M_1}.
\ea

Now we consider a space $M_n$, which is defined by  left picture of (c) in Fig.\ref{replicafig}.
This replicated geometry has a angle $2\pi n-\theta_A$ wedge along $\Gamma$. The gravity action 
on this space is 
\ba
I_{M_n}=-\frac{1}{16\pi G_N}\int_{M_n} \s{g}(R-2\Lambda)-\frac{1}{8\pi G_N}\int_{\de M_n} \s{h}K
+\frac{1}{8\pi G_N}\int_{\Gamma}\left((2n-1)\pi-\theta_A\right)\s{\gamma}. \no \label{mnac}.
\ea
Thus if we past the space $M_A$ to $M_n$, which is called $R_n$, we get the full replicated space without any boundaries. This is the replicated geometry for the standard calculation of entanglement entropy. The gravity action on $R_n$ eventually reads
\ba
I_{R_n}=I_{M_n}+I_{M_A}=-\frac{n}{16\pi G_N}\int_{R^{d+1}} \s{g}(R-2\Lambda)
+\frac{1}{8\pi G_N}\int_{\Gamma}(2n-2)\pi\s{\gamma},  \label{mrac}
\ea
where notice the equivalence $R_1=R^{d+1}$.

Thus the entanglement entropy $S_{\Sigma_A}$ is found to be 
\ba
S_{\Sigma_A}&=&-\frac{\de}{\de n} \log \frac{Z_{R_N}}{(Z_{R^{d+1}})^n}\Biggr |_{n=1}\no
&=& \frac{A(\Gamma)}{4G_N}.  \label{areaent}
\ea

The above argument is quite general and can be applied to any classical gravity.
In particular we can obtain the holographic entanglement entropy in AdS/CFT \cite{RT}
from the above method using the Hayward term, which is equivalent to the argument \cite{Fursaev:2006ih}.
Also our argument can be applied to a more general conjecture for any spacial surfaces \cite{Bianchi:2012ev}.
It is clear from the above calculation that the Hayward term is crucial to obtain the gravitational entropy.

\subsection{Pasting Rules of Gravitational Spaces}

Here we would like to explore the general rules of pasting gravitational spaces where the Hayward term plays the important role.
Our argument in this subsection is a reformulation of the much earlier work \cite{Brill:1994mb}.

\begin{figure}
  \centering
  \includegraphics[width=10cm]{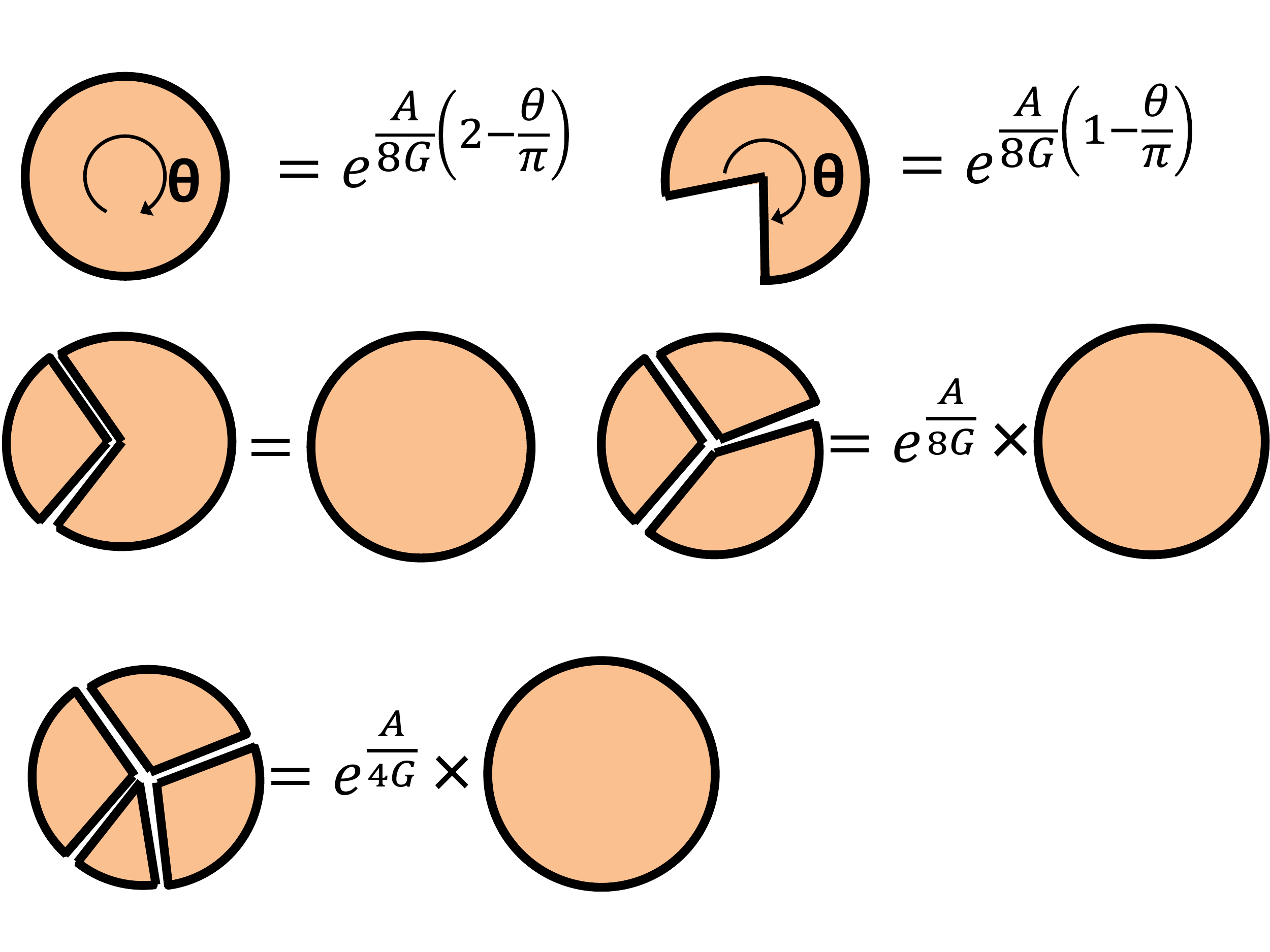}
  \caption{Pasting Rules of Gravity Action.}
\label{pastefig}
\end{figure}

Let us start with summarizing the previous replica computation. 
The wedge geometry with the angle $\theta$ leads to the Hayward term contribution to a gravity partition function $Z_{W(\theta)}$ on the wedge
\ba
Z_{W(\theta)}\propto e^{\frac{A(\Gamma)}{8G_N}\left(1-\frac{\theta}{\pi}\right)},  \label{wedgezz}
\ea
as depicted in the top right picture in Fig.\ref{pastefig}.  Notice that the terms other than Hayward term is additive and we omitted in the above.
We attached another wedge with an angle $\theta'$ to the previous wedge and get a geometry with the angle $\theta+\theta'$,
which does not have any boundaries. This partition function is given by the multiplications of the two:
\ba
Z_{W(\theta)}\cdot Z_{W(\theta')}\propto  e^{\frac{A(\Gamma)}{8G_N}\left(2-\frac{\theta+\theta'}{\pi}\right)}=Z_{C(\theta+\theta')}.
\ea
This coincides with the well-known result of gravity partition function $Z_{C}$ with the conical geometry, whose deficit angle $\delta$ is given 
by $\delta=2\pi-\theta-\theta'$:
\ba
Z_{C(2\pi-\delta)}\propto e^{\frac{A(\Gamma)}{4G_N}\delta}.  \label{deficit}
\ea
This simply follows from the familiar fomula $R=4\pi\delta\cdot \delta(\Gamma)$.

In this way, when we paste two spaces along their boundaries, the resulting gravity partition function agrees with that of the connected geometry.
However, this is not true when we paste more than two spaces as depicted in the final two pictures in  Fig.\ref{pastefig}.  When we paste $k$ wedge spaces 
with the angle $\theta_1$, $\theta_2$,.. $\theta_k$, we have 
\ba
\prod_{i=1}^{k}Z_{W(\theta_i)}\propto  e^{\frac{A(\Gamma)}{8G_N}\left(k-\sum_{i=1}\frac{\theta_i}{\pi}\right)}=e^{\frac{A(\Gamma)}{8G_N}(k-2)}\cdot 
Z_{C(\sum_i\theta_i)}. \label{pastf}
\ea

\begin{figure}
  \centering
  \includegraphics[width=10cm]{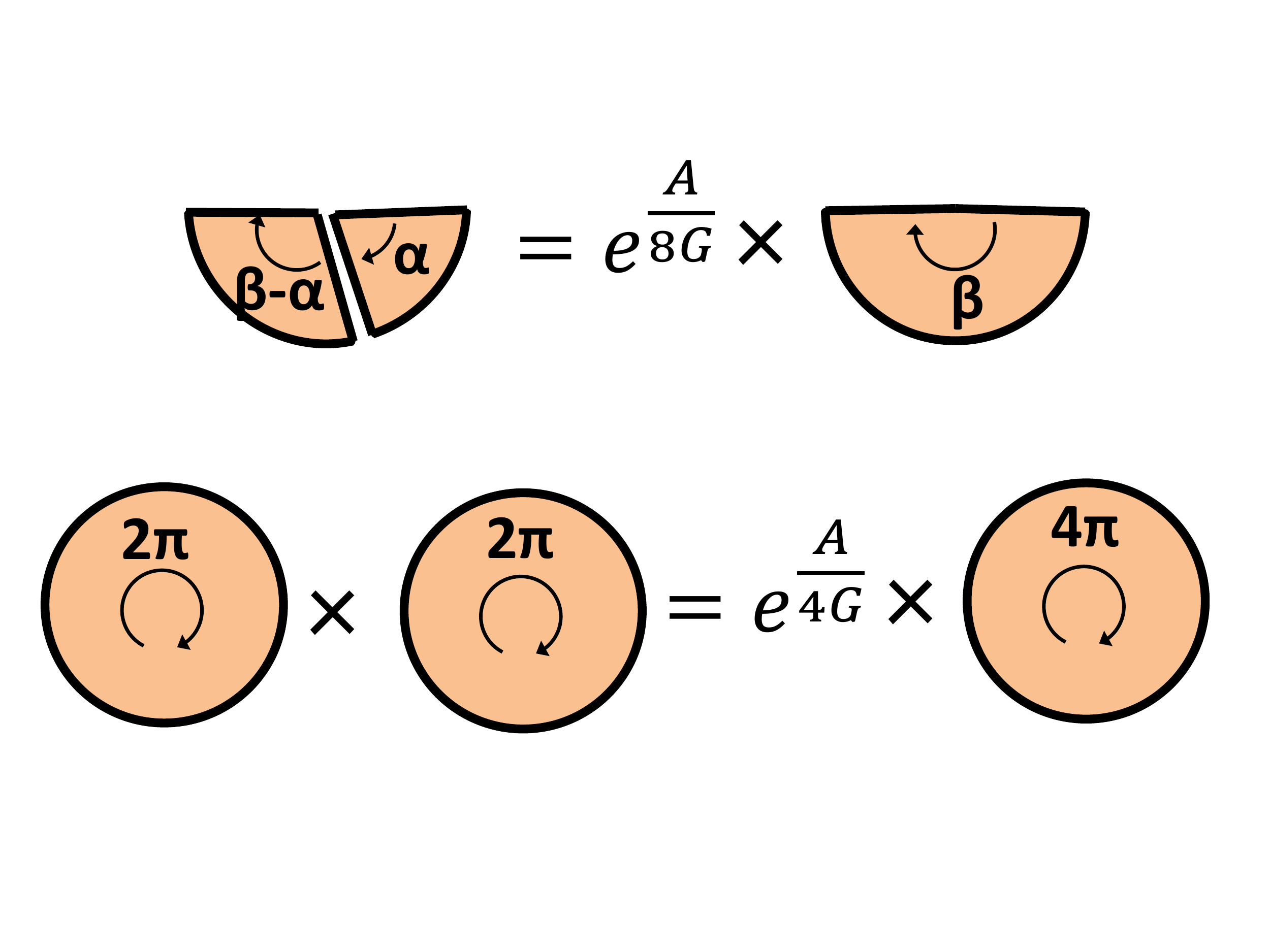}
  \caption{Pasting Two Wave Functions (Top) and its Doubled Version (Bottom). }
\label{pastefigg}
\end{figure}

We can also turn to the pasting of two wedges into a larger wedge as depicted in the upper picture of Fig.\ref{pastefigg}:
\ba
Z_{W(\theta_1)}\cdot Z_{W(\theta_2)}=e^{\frac{A(\Gamma)}{8G_N}}\cdot Z_{W(\theta_1+\theta_2)}.  \label{pwf}
\ea
We can view this as a pasting rule of Hartle-Hawking wave function.

By taking two copies of this, we can estimate the partition function in the presence of conical angle:
\ba
Z_{C(2\theta_1)}\cdot Z_{C(2\theta_2 )}=e^{\frac{A(\Gamma)}{4G_N}}\cdot Z_{C(2\theta_1+2\theta_2)},
\ea
which reproduces (\ref{deficit}).

\section{Hayward Term and CFT Duals}\label{sec:hayward_cft}

As we have explained, the pasting rule (\ref{pwf}), which gives an anomalous boundary contribution, can be regarded as the basic 
rule from which we can derive other pasting rules including the conical angle contributions. Therefore here we would like to
consider a holographic interpretation of  (\ref{pwf}) in the light of AdS/CFT. 
 
\begin{figure}
  \centering
  \includegraphics[width=6cm]{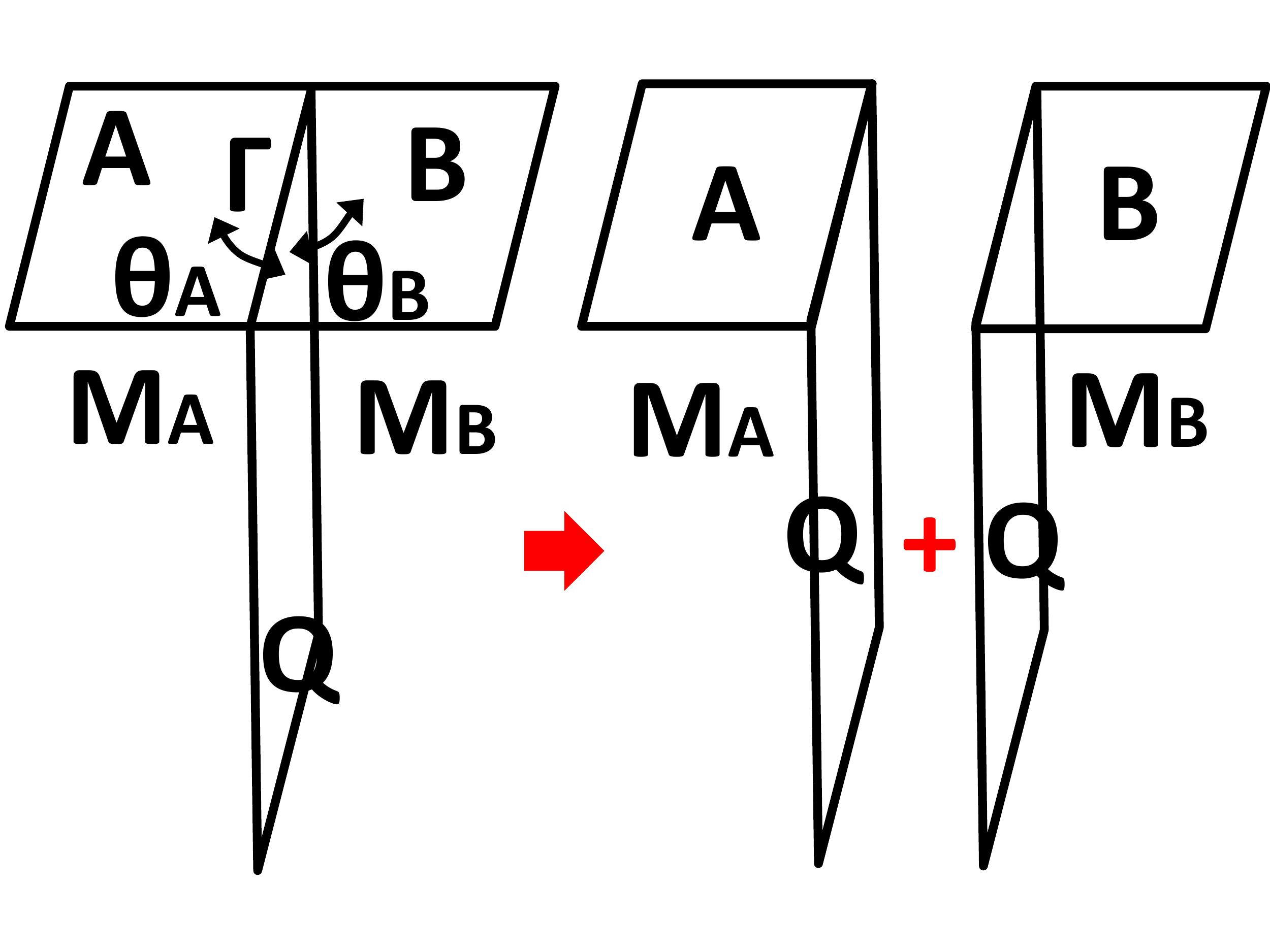}
  \caption{Separating a Poincare AdS into Two Parts $M_A$ and $M_B$. The boundary of the Poincare 
AdS is $A\cup B$, which is a plane. The separation is done along the surface $Q$.}
\label{holfig}
\end{figure}

We start with the gravity on a Poincare AdS$_{d+1}$ (with the radius $R_{AdS}$), 
\ba
ds^2=R_{AdS}^2\left(\frac{dz^2+\sum_{a=1}^d dx^a dx^a}{z^2}\right),
\ea
dual to a $d$ dimensional holographic CFT on $R^d$ as depicted in the left of Fig.\ref{holfig}. As usual in AdS/CFT we set the boundary at $z=\ep$ and 
regard $\ep$ as the UV cut off scale of the dual CFT. 
We introduce a surface $Q$ so that the Poincare AdS is divided into two parts $M_A$ and $M_B$ as in the right picture of Fig.\ref{holfig}. 
The AdS boundary is divided into $A$ and $B$.  The angle between $A$ and $Q$ is called $\theta_A$ and then that between $Q$ and $B$ is given by
$\theta_B=\pi-\theta_A$.

\subsection{Generalized Holography Viewpoint (Dirichlet-Dirichlet Case)}\label{subsec:wf}

Consider a holographic dual of $M_A$ and $M_B$, separately. We impose the Dirichlet boundary condition on both $M_A$, $M_B$ and $Q$ in 
the gravity side.  Then, since $\de M_A=A\cup Q$, we expect that the gravity on $M_A$ is dual to a CFT on $A\cup Q$. 
However, $Q$ is not an asymptotically AdS boundary, the theory on $Q$ should be a non-local field theory, which is obtained from the original CFT 
via the inhomogeneous RG flow down to the cut off scale $\ep_Q=O(1)$. 
On the other hand, the field theory on $A$ is the original CFT with the UV cut off scale $\ep$.   Notice that this type of generalized 
holography follows from the surface/state correspondence  \cite{Miyaji:2015yva} and if we set $\theta_A=\pi/2$ this coincides with a 
specific example of that appears in holographic entanglement of purification \cite{UT,Nguyen:2017yqw}.
We define $c$ to be the central charge of the 
holographic CFT such that $c\propto \frac{R^{d-1}}{G_N}$.
The partition function on $A\cup Q$ can be estimated as 
\be
Z_{AQ}= Z_{A}\cdot Z_{Q},
\ee
where 
\be
Z_A= e^{\lambda_A\cdot cV(A)},\ \ \ \ Z_Q= e^{(d-1)\lambda_Q\cdot cV(Q)},
\ee
where $\lambda_A$ and $\lambda_Q$ are O(1) numerical constants. 
Here $V(A)$ and $V(Q)$ are the volume of $A$ and $Q$, which look like
\ba
&& V(A)=\int_A \frac{dx^d}{z^d}=\frac{A(\Gamma)\cdot L_A}{\ep^d},\no
&& V(Q)=\int_Q \frac{dx^{d-1}dz}{z^d}= \frac{A(\Gamma)}{(d-1)\ep^{d-1}},
\ea
where $L_A=\infty$ in the horizontal length of half-plane $A$. This leads to the estimation:
\ba
Z_{Q}= e^{\lambda_Q\cdot c\cdot  \frac{A(\Gamma)}{\ep^{d-1}}}. \label{zedge}
\ea

On the other hand, from the gravity side, the pasting rule (\ref{pwf}) and the bulk-boundary duality in AdS/CFT leads to
\ba
Z_{AQ}\cdot Z_{BQ}=Z_{AB}\cdot e^{\frac{A(\Gamma)}{8G_N}}.  \label{sumrt}
\ea
Note that since $\theta_A+\theta_B=\pi$, the Gibbons-Hawking contribution cancel with each other and the terms other than Hayward term 
simply get additive. 

In addition, the following equality is obvious 
\ba
Z_{A}\cdot Z_{B}=Z_{AB}.
\ea
Therefore the extra contribution in (\ref{sumrt}) is qualitatively explained by the contribution (\ref{zedge}) from the (non-local) theory on $Q$.

\begin{figure}
  \centering
  \includegraphics[width=6cm]{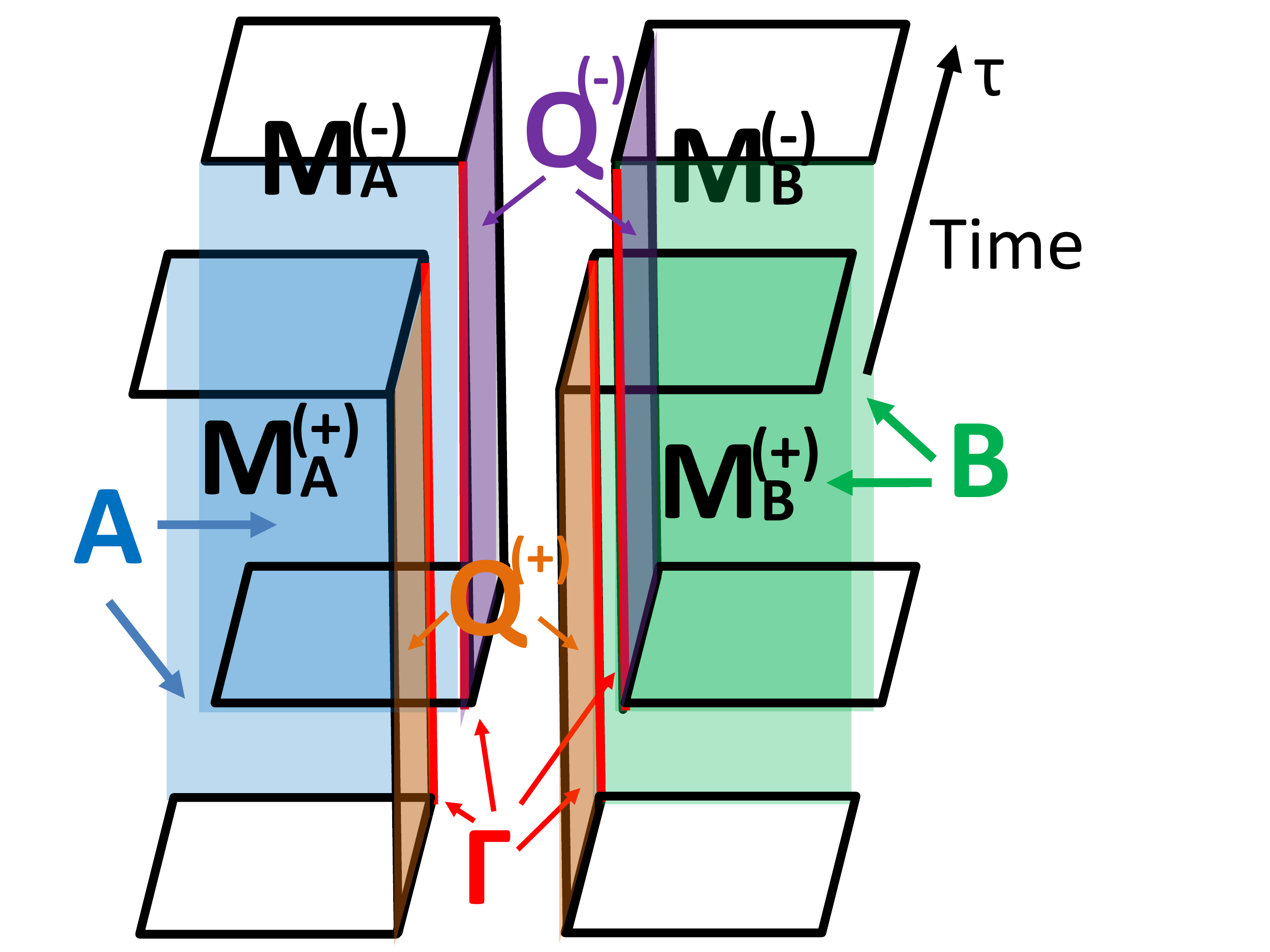}
  \caption{Separating a space into four subspaces $M^{(+)}_A, M^{(-)}_A, M^{(+)}_B$ and $M^{(-)}_B$, 
which are extended in the upper and lower directions without encountering any boundaries.
The boundary of $M^{(\pm)}_A$  (and $M^{(\pm)}_B$)  consists of $A\cup Q^{(\pm)}$ 
(and  $B\cup Q^{(\pm)}$).  The four subspaces have a common surface $\Gamma$.}
\label{holfiwg}
\end{figure}

Next we would like to turn to the viewpoint of wave functionals in 
a $d+1$ dimensional gravitational theory.  
For this, we start with a spacetime $M$ with no boundaries and divide it into four subspaces, named
$M^{(+)}_A, M^{(-)}_A, M^{(+)}_B$ and $M^{(-)}_B$, as in Fig.\ref{holfiwg}. The separation is done 
with respect to the $d$ dimensional surfaces $A$, $B$, $Q^{(+)}$ and $Q^{(-)}$, such that 
$\de M^{(\pm)}_A=A\cup Q^{(\pm)}$ and  $\de M^{(\pm)}_B=B\cup Q^{(\pm)}$. We define the $d-1$ dimensional surface $\Gamma$ as the intersection between $A$ and $M^{(\pm)}_A$ (or equally  $B$ and $M^{(\pm)}_B$). Note that the four subspaces share the same codimension two surface $\Gamma$. 

The gravity partition function on $M^{(+)}_{A}$ and $M^{(+)}_B$, denoted by 
$Z_{M^{(+)}_{A}}$ and $Z_{M^{(+)}_{B}}$ are interpreted as the 
(un-normalized) gravity wave functionals:
\ba
|\Psi_{A}\lb=\sum_{i=1}^{N} |i\lb_A |i\lb_{Q_A},\ \ \ |\Psi_{B}\lb=\sum_{j=1}^N |j\lb_B |j\lb_{Q_B}. \label{eq:wf}
\ea
Here note that the state $|\Psi_{A}\lb$ lives in the Hilbert space which comes from the configurations on $A$ and those on $Q^{(+)}$, whose bases are denoted as $|i\lb_A$ and  $|i\lb_{Q_A}$ (the same comment is applicable to $|\Psi_{B}\lb$).

Since these provide the purifications of the density matrix $\rho_A$ and $\rho_B$, 
$N$ is given by the exponential of the entanglement entropy $S_A=S_B$ between $A$ and $B$
\ba
N=e^{S_A}=e^{\frac{A(\Gamma)}{4G_N}},
\ea
where we employed the conjecture in \cite{Bianchi:2012ev} for the identification 
$S_A=\frac{A(\Gamma)}{4G_N}$ in gravity. 
In the same way, the gravity path-integrals on  $M^{(-)}_{A}$ and $M^{(-)}_{B}$ are dual to 
$\la \Psi_A|$ and $\la \Psi_B|$, respectively.

Their inner product is estimated as
\ba
(\la \Psi_A|\la \Psi_B|)\cdot (|\Psi_{A}\lb|\Psi_{B}\lb)=N^2,
\ea
which is expected to be dual to the multiplication of partition functions:
\ba
Z_{M^{(+)}_{A}}\cdot Z_{M^{(+)}_{B}}\cdot Z_{M^{(-)}_{A}}\cdot Z_{M^{(-)}_{B}}.
\ea

On the other hand, $Z_{M^{(+)}_{A}\cup M^{(+)}_{B}}$ is dual to the quantum state
\ba
|\Psi_{AB}\lb=\sum_{i=1}^{N} |i\lb_A |i\lb_B.
\ea
Similarly, $Z_{M^{(-)}_{A}\cup M^{(-)}_{B} }$ is dual to $\la \Psi_{AB}|$.
Their inner product is given by
\ba
\la \Psi_{AB}|\Psi_{AB}\lb=N,
\ea
and this corresponds to the full partition function 
$Z_{M^{(+)}_{A}\cup M^{(+)}_{B}\cup M^{(-)}_{A}\cup M^{(-)}_{B}}$.

In this way, our CFT interpretation predicts  
\ba
Z_{M^{(+)}_{A}}\cdot Z_{M^{(+)}_{B}}\cdot Z_{M^{(-)}_{A}}\cdot Z_{M^{(-)}_{B}}
=e^{\frac{A(\Gamma)}{4G_N}}\cdot Z_{M^{(+)}_{A}
\cup M^{(+)}_{B}\cup M^{(-)}_{A}\cup M^{(-)}_{B}}.
\ea
This reproduces the pasting rule (\ref{pastf}).

\subsection{AdS/BCFT Viewpoint (Dirichlet-Neumann Case)}\label{subsec:bcft}

If we impose the Neumann boundary condition on $Q$ in the previous setup of Fig.\ref{holfig}
and the Dirichlet boundary condition on both $A$ and $B$, we can regard this as a setup of AdS/BCFT \cite{AdSBCFT}.  
In this case, the gravity on $M_A$ (or $M_B)$ is dual to the CFT on $A$ with an appropriate boundary condition imposed on $Q$.
Below in this subsection we would like to give an interpretation of the area entropy $A(\Gamma)/4G_N$ in gravity in terms of the CFT on a manifold with 
a boundary.

The AdS/BCFT argues that the CFT partition function on $A$ with a certain boundary condition 
labeled by $i$ imposed on  $\de A=\Gamma$, denoted by $Z^{BCFT(i)}_A$,
 is equal to the gravity partition function on $M_A$ with a suitable boundary condition (again labeled by $i$) 
imposed on the surface $Q$:
\ba
Z^{BCFT(i)}_A=e^{-\ti{I}_{M_A(i)}}|_{on-shell},   \label{gravityma}
\ea
where the gravity action $\ti{I}_{M_A(i)}$ is defined by
\ba
\ti{I}_{M_A(i)}=-\frac{1}{16\pi G_N}\int_{M_A} \s{g}(R-2\Lambda)-\frac{1}{8\pi G_N}\int_{A} \s{h}K
-\frac{1}{8\pi G_N}\int_{Q} \s{h}(K+L^{(i)}_m)+I_{H} \no
\ea
where $L^{(i)}_m$ is the matter Lagrangian localized on the surface $Q$ and $i$ labels all possible boundary conditions of the CFT; the term $I_{H}$ denotes the 
Hayward term.

The variational principle leads to the Neumann-type boundary condition on $Q$
\ba
K_{ab}(x)-K(x)h_{ab}-T^{(i)}_{ab}(x)=0,  \label{bein}
\ea
where $T^{(i)}_{ab}$ is the boundary energy stress tensor from the matter described by the Lagrangian 
$L^{(i)}_m$. In particular, when it takes the special form $T^{(i)}_{ab}=-T_B h_{ab}$, the boundary condition preserves a part of conformal symmetry and 
the gravity dual is called the boundary conformal field theory (BCFT). The parameter $T_B$ represents the tension when we regard the surface $Q$ as a 
end of world brane and the solving the boundary conterpart of Einstein equation (\ref{bein}) uniquely chooses the angle $\theta_A$ as a function of $T_B$ \cite{AdSBCFT}.
However, in our arguments here we allow all possible boundary conditions which are not conformally invariant in general. 

Now, in the CFT, we can decompose the partition function as 
\ba
Z_{AB}=\sum_i \la 0|e^{-L_A H/2}|i\lb \la i|e^{-L_B H/2}|0\lb,
\ea
where $L_{A,B}$ is the horizontal length of $A$ and $B$;  $|i\lb$ labels all states in the CFT. We regard the horizontal direction in Fig.\ref{holfig} as the Euclidean time and 
the Hamiltonian in this time evolution is denoted by $H$. 

The gravity dual of $\la 0|e^{-L_A H/2}|i\lb$ is given by the gravity partiton function $Z^{BCFT(i)}_A$ in the AdS/BCFT as explained just before.
In general it describes a geometry in a region surrounded by $A$ and $Q$ as in Fig.\ref{holfig}.  The profile of $Q$ depends on the boundary condition 
specified by $|i\lb$ and the angle $\theta_A$ also depends on this boundary condition. The number of boundary conditions $N_B$ is the same as the number of states 
when we regard the horizontal direction as the Euclidean time, which is estimated by 
\ba
N_B\sim e^{c\frac{A(\Gamma)}{\ep^{d-1}}},  \label{nbct}
\ea
up to an $O(1)$ undetermied coefficient.  This estimation obeys from the standard formula of the entropy in CFTs at finite temperature: $S\sim cA(\Gamma)T^{d-1}$ 
 by regarding the cut off energy $1/\ep$ as the effective temperature $T$.   This counting (\ref{nbct}) of the number of all possible boundary conditions agrees with the gravitational entropy (\ref{areaent}) up to an $O(1)$ factor. This provides another interpretation of the area entropy in the light of AdS/BCFT.

\subsection{Hayward Term from AdS/BCFT}\label{HayBCFT}

\begin{figure}
  \centering
  \includegraphics[width=10cm]{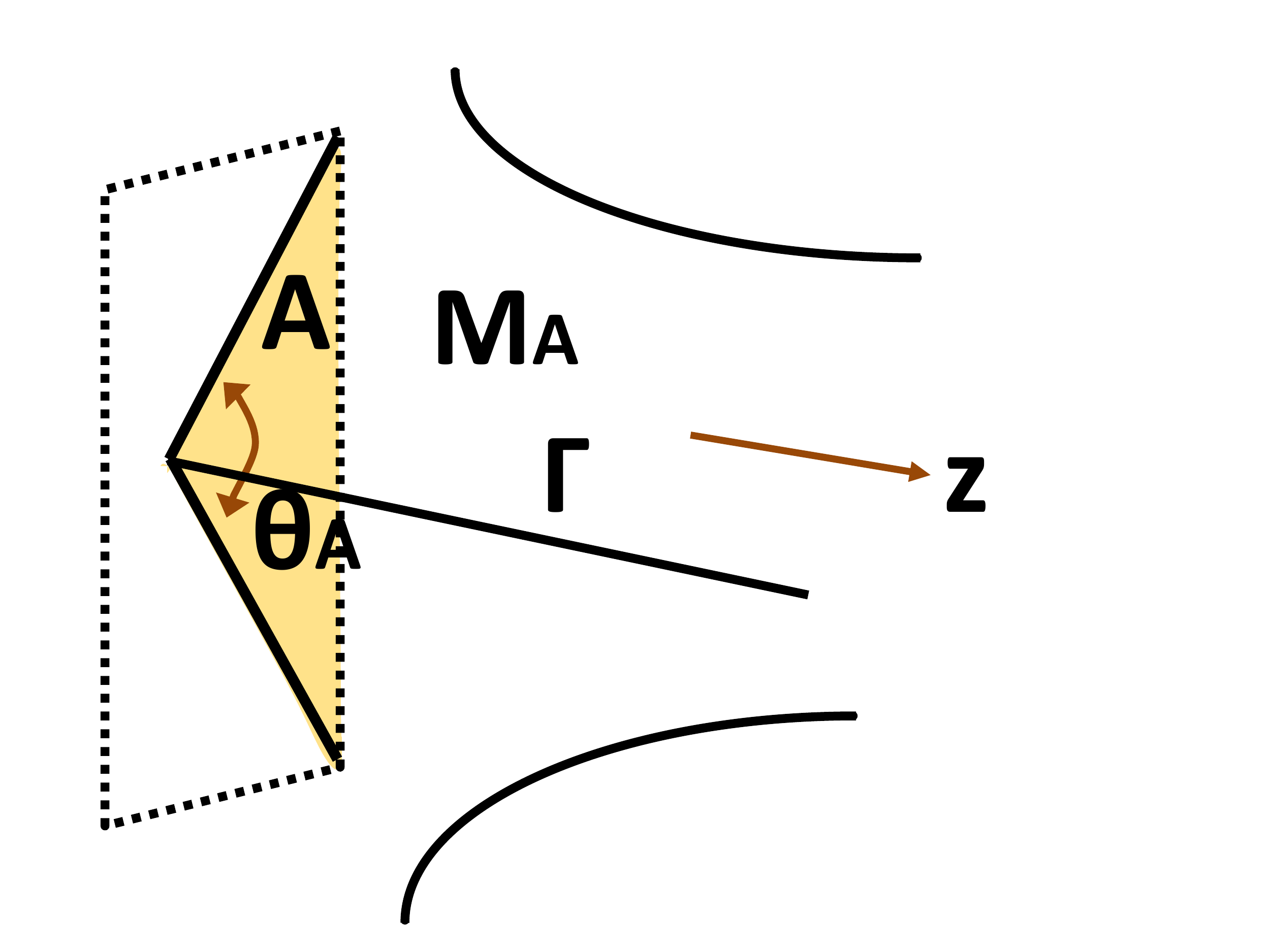}
  \caption{A gravity dual of a CFT on a wedge in AdS/BCFT. }
\label{figwedge}
\end{figure}

In a class of  AdS/BCFT setups where the wedges exist in the AdS boundaries, 
 we can explicitly confirm the necessity of Hayward term in the bulk wedge. Below we would like to explain this.
Consider a two dimensional CFT on a wedge region $A$ with the angle $\theta_A(>0)$:
\ba
A=\{(x,y)|\ |y|<\tan\frac{\theta_A}{2}\cdot x,\ x>0\}.
\ea
We impose a conformally invariant boundary condition on the boundary $\de A$ and apply the AdS/BCFT. 
Its gravity dual $M_A$ has also a wedge and its corner is denoted by $\Gamma$ as depicted in Fig.\ref{figwedge}.
We impose a Neumann-like boundary condition (\ref{bein}) on the boundary $Q$ which extends in the bulk.
Even though in general we need to take into account back-reactions due to the brane $Q$ by solving the Einstein equation with the 
boundary condition (\ref{bein}), the geometry near the asymptotically AdS region is not changed. Therefore the divergent contribution to the 
holographic partition function for $A$ is universally estimated in the presence of Hayward term as follows
\ba
Z^{gravity}_A\sim e^{\# \ep^{-2}+\frac{c}{12}\left(1-\frac{\theta_A}{\pi}\right)\log\ep^{-1}},  \label{gravitzdgc}
\ea
where $\ep$ is the UV cut off in 2d CFT.
Here the quadratically divergent term comes from the bulk integral of Einstein-Hilbert action on $M_A$, while the logarithmic one comes from the Hayward term
as in (\ref{wedgezz}):
\ba
\frac{A(\Gamma)}{8G_N}\left(1-\frac{\theta_A}{\pi}\right)=\frac{1}{8G_N}\int_\ep \frac{dz}{z} \left(1-\frac{\theta_A}{\pi}\right)
=\frac{c}{12}\left(1-\frac{\theta_A}{\pi}\right)\log\ep^{-1}. 
\ea

On the other hand, in the CFT side, the standard argument of conformal anomaly fixes the coefficient of the logarithmic term in terms of Euler number $\chi(A)$ of 
the manifold $A$ as follows:
\ba
Z^{CFT}_A\sim e^{\frac{c}{6}\chi(A)\log \ep^{-1}}.  \label{spherepe}
\ea
For the wedge geometry $A$, the Euler number is computed as 
\ba
\chi(A)=\frac{1}{4\pi}\int_A \s{g}R+\frac{1}{2\pi}\int_{\de A} K\s{\gamma}+\frac{1}{2\pi}(\pi-\theta_A),
\ea
where the last term is the ``Hayward term'' in the Euler number and is derived by regularizing 
the boundary Gibbons-Hawking term. Since we have $R=0$ on $A$ and $K=0$ on $\de A$, this leads to 
\ba
Z^{CFT}_A\sim e^{\frac{c}{12} \left(1-\frac{\theta_A}{\pi}\right)\log \ep^{-1}},
\ea
which agrees with the gravity result (\ref{gravitzdgc}).

In this way, the Hayward term on $\Gamma$ in the gravity is necessary to reproduce the correct conformal anomaly 
in the dual CFT.

\section{Comparison with Edge Modes in Gauge Theories}

In section \ref{sec:hayward_cft}, we have introduced two AdS/CFT realizations of the gravitational subsystems, namely the generalized holography viewpoint (section \ref{subsec:wf}) and the AdS/BCFT viewpoint (section \ref{subsec:bcft}). Here we point out there are nice counterparts in gauge theories\footnote{These are no more than analogies. Do not confuse these with ones in the real gauge theories dual to gravity. There are probably no direct correspondence of edge modes themselves in the gauge/gravity duality.}. 
The existence of edge modes has been discussed extensively in the context of entanglement entropy in gauge theories. Refer to \cite{Buividovich:2008gq,Donnelly:2011hn,Ghosh:2015iwa,Aoki:2015bsa,Eling:2013aqa,Agon:2013iva,Gromov:2014kia,Radicevic:2014kqa,Donnelly:2014gva,Casini:2014aia,Huang:2014pfa,Donnelly:2014fua,Donnelly:2015hxa,Soni:2015yga,VanAcoleyen:2015ccp,Zuo:2016knh,Donnelly:2016mlc,Dowker:2017flz,Hung:2015fla,Radicevic:2015sza,Ma:2015xes,Itou:2015cyu,Casini:2015dsg,Radicevic:2016tlt,Nozaki:2016mcy,Soni:2016ogt,Aoki:2016lma,Balasubramanian:2016xho,Agarwal:2016cir,Aoki:2017ntc,Hategan:2017jts,Hung:2017jnh,Pretko:2018nsz,Blommaert:2018rsf,Hategan:2018uas,Anber:2018ohz,Blommaert:2018oue,Lin:2018bud,Huerta:2018xvl,Moitra:2018lxn,Casini:2019nmu,Belin:2019mlt} for recent developments. 
As similar to the diffeomorphism in gravity, we need to manage the gauge degrees of freedom at the boundary with great care. One way to define the subsystem is to embed the gauge-invariant Hilbert space into a larger one, so-called ``extended Hilbert space''\cite{Buividovich:2008gq,Donnelly:2011hn,Ghosh:2015iwa,Aoki:2015bsa}. There are in principle many choices for implementing the enlarged Hilbert space. We list two of them, which are relevant to the previous AdS/CFT examples. 
\begin{enumerate}
\item Add the dynamical degrees of freedom at the boundary so that one can make each subsystem gauge-invariant. It can be also interpreted as the ``purification'' of the classical correlations due to the gauge constraints. This will be discussed in section  \ref{subsec:purification}  below and is analogous to section \ref{subsec:wf}. 
\item Start from the Hilbert space which includes gauge-variant states. Such states will be projected out when we evaluate the gauge-invariant states. This will be discussed in section \ref{subsec:ext} below  and is analogous to section  \ref{subsec:bcft}. 
\end{enumerate}

In any case, we need to introduce additional degrees of freedom, especially at the boundary --- so-called edge modes. Then, the entanglement entropy for the gauge theories has the following form:
\ba\label{eq:gaugeEE}
S(\rho_A)=-\sum_jp_j\log p_j +\sum_j p_j \sum_a\log d^{(j)}_a +\sum_j p_jS(\rho^{(j)}_A),
\ea
where the final term is the standard entanglement entropy which is distillable. In what follows, we spell out the remaining parts which come from the correlations of edge modes. The first term comes from the Shannon entropy due to the probability $p_j$ associated with the superselection sectors, labeled by $j$. Each $j$ is determined by the expectation values for the Casimir operators (boundary electric fluxes) at the boundary points. The second term counts the correlations from the color degrees of freedom, which may be regarded as a sum of the expectation value of local operators. Here $d^{(j)}_a$ represents the dimension of an irreducible representation of the color flux at each point $a$. 

One can expect we have the similar form as \eqref{eq:gaugeEE} even for the generalized entropy in gravity. In this case, the label $j$ would be determined from the eigenvalues of the area operator, which would play a role of the Casimir operator\cite{Donnelly:2016auv,Harlow:2016vwg} (see also recent works in this direction\cite{Lin:2017uzr,Akers:2018fow,Dong:2018seb,Dong:2019piw,Lin:2018xkj,Jafferis:2019wkd}). In particular, we will discuss the counterpart of the second term in \eqref{eq:gaugeEE} since we have focused on the area-fixed sector. 

\begin{figure}
  \centering
  Gauge theory (Wilson line) setup
  \includegraphics[width=15cm]{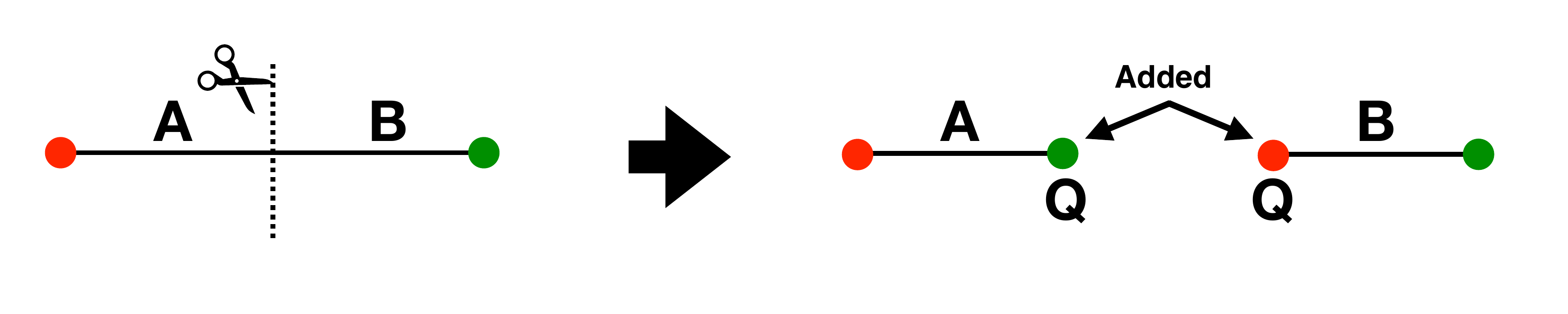}
  Gravity (AdS/CFT) setup
    \includegraphics[width=15cm]{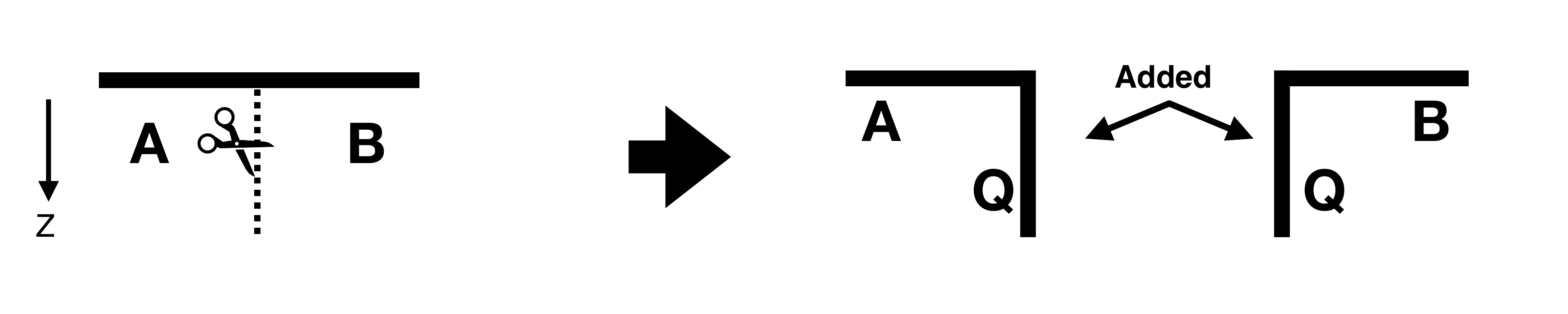}
  \caption{Purification approach to define the subsystem $A$ and $B$. The edge modes (states living on $Q$) play a role of ``purifiers'' due to the correlations from gauge invariance. In the bottom figure, the labels $A, B$ should be regarded as $A_-,B_-$ in Figure \ref{holfiwg}. }
\label{wl1}
\end{figure}

\subsection{Doubling the Degrees of Freedom}\label{subsec:purification}
A simple and analogous example of the area-fixed state in gauge theories is a meson or Wilson line state across the boundary\footnote{The similar state in QED is discussed in \cite{Harlow:2015lma}. A lattice counterpart of such meson states are discussed extensively in \cite{Aoki:2017ntc}.}. 
If we naively split it, the gauge transformation at the boundary makes the state different one. A simple rescue to cure the gauge invariance in the subsystem is to add infinitely heavy charged degrees of freedom at the boundary. See the upper panel of figure \ref{wl1}. Then, one may view the state in subsystem $A$, say $\rho_A$, as
\ba
\rho_A=\dfrac{1}{N_G}\sum^{N_{G}}_{n=1}\ket{n}_A\bra{n}_A,
\ea
where we assumed the state belongs to the fundamental representation of $SU(N)$ gauge group for concreteness. Each $n$ represents color degrees of freedom. The form should be uniquely fixed up to trivial phase factors since the state must be gauge-singlet. We can interpret it as a purification of the classical correlations due to the gauge constraints:
\ba\label{eq:purification_gauge}
\rho_A\longrightarrow\ket{\psi}_{AQ}=\dfrac{1}{\sqrt{N_G}}\sum^{N_G}_{n=1}\ket{n}_A\ket{\bar{n}}_{Q}.   \label{wqw}
\ea
The degrees of freedom at $Q$ corresponds to the aforementioned heavy charged states, namely the edge modes. 

In our AdS/CFT setup (section \ref{subsec:wf}), we can view each $\bar{n}$ in \eqref{eq:purification_gauge} as new degrees of freedom living on a new ``boundary''. See lower panel of figure \ref{wl1}.   
This contribution is analogous to the second term with fixed $j$ in \eqref{eq:gaugeEE}.   The state (\ref{wqw}) is analogous to the state (\ref{eq:wf}).

We know that the holographic Renyi entropy does not admit the flat spectrum. 
It means that the usual minimal area surface cannot be regarded as an edge mode living on a single super-selection sector. Our argument so far has been basically the area-fixed argument \cite{Akers:2018fow,Dong:2018seb}, namely we have focused on a single sector and neglected distillable contributions (bulk matter fields). We stress that our interpretation at the semi-classical limit will hold even after inclusions of matter fields and so on. This is simply because classical and distillable parts are decoupled as in \eqref{eq:gaugeEE}. 

From the macroscopic viewpoint, the area-fixing can be understood as a change of the statistics. In this sense, the addition of Hayward term is just the Legendre transformation.

\subsection{Including the Gauge-variant States}\label{subsec:ext}

\begin{figure}
  \centering
  Gauge theory (Wilson line) setup
  \includegraphics[width=15cm]{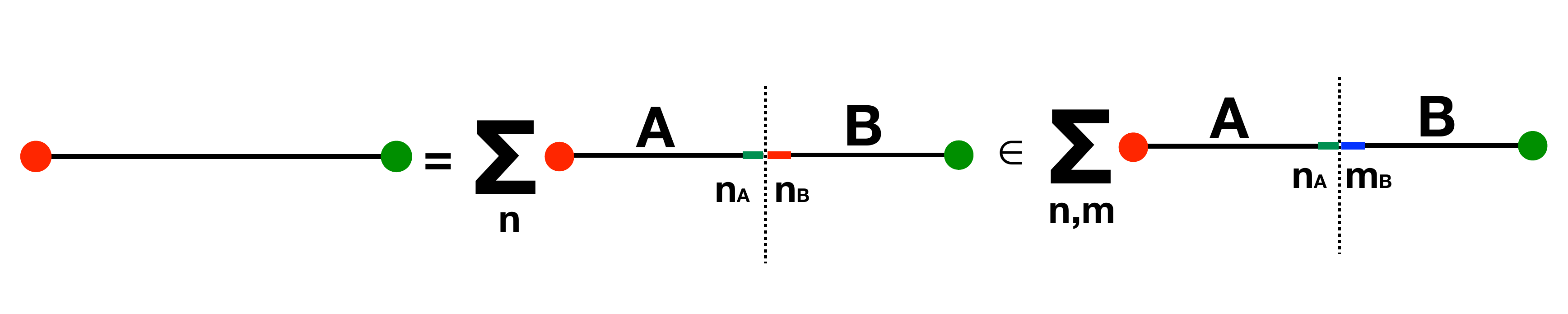}\vspace{4mm}
  Gravity (AdS/BCFT) setup\vspace{4mm}
    \includegraphics[width=15cm]{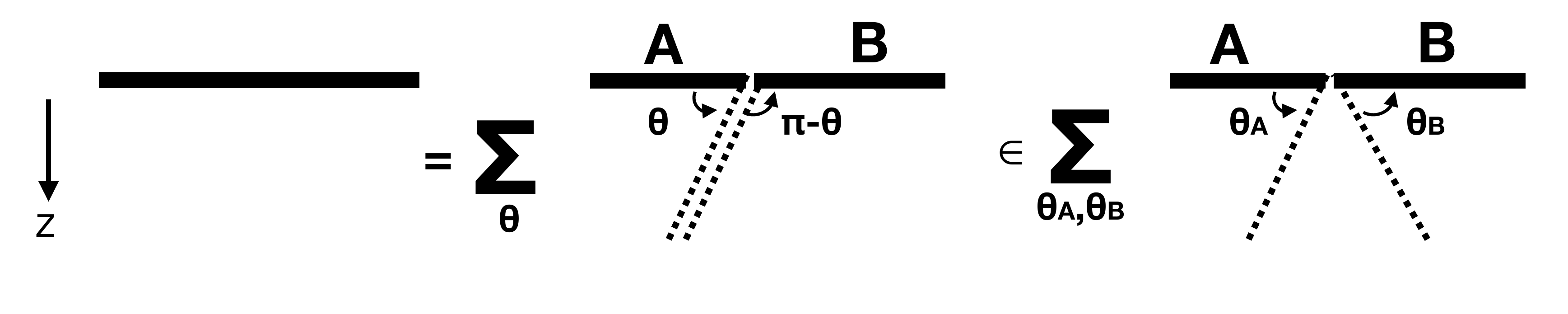}
  \caption{Schematic pictures on embedding the physical states into the extended Hilbert space includes all gauge-variant states in gauge theories and gravity (AdS/BCFT).}
\label{wl2}
\end{figure}

One can also extend the gauge-invariant Hilbert space $\mathcal{H}_{ginv.}$ such that it includes the gauge-variant states in $\mathcal{H}^{\perp}_{ginv.}$. In the previous example for the meson or Wilson line state, let us say $\ket{\psi}$, it means that
\ba
\ket{\psi}=\sum^{N_G}_{n=1}\ket{n}_A\ket{\bar{n}}_B\in\mathcal{H}_{ginv.}\subset\mathcal{H}_{ext.}\equiv\mathcal{H}_{ginv.}\oplus\mathcal{H}^{\perp}_{ginv.}, \label{eq:ext_phys}
\ea
where $\mathcal{H}_{ext.}$ allows the state such as $\sum^N_{n,m=1}c_{n,m}\ket{n}_A\ket{\bar{m}}_B$. 
In the light of the extended Hilbert space $\mathcal{H}_{ext.}$, we can safely say $\mathcal{H}_{ext.}$ is bi-partite with respect to any subsystems. 
Then, we have ``entanglement'' from color degrees of freedom which are protected by gauge-invariance, so not distillable. See upper panel of figure \ref{wl2}.

Our AdS/BCFT setup in section \ref{subsec:bcft} is quite similar to the state \eqref{eq:ext_phys} in the gauge theories. This is because we implicitly allowed our Hilbert space to be spanned by $\ket{\theta_A}\ket{\theta_B}$ with $\theta_A+\theta_B\neq\pi$. The generic states with $\theta_A+\theta_B\neq\pi$ do not satisfy the constraints from diffeomorphism invariance. Hence, such states are out of gauge-invariant Hilbert space (Lower panel of figure \ref{wl2}). The ``physical'' states must belong to the subspace with $\theta_A+\theta_B=\pi$. Now the number of color degrees of freedom $N_G$ corresponds to the number of all possible choices of the boundary energy momentum tensor of the boundary surface in the gravity side, which is dual to the boundary condition (\ref{bein}) in AdS/BCFT.

\section{Edge Modes as Open Strings}

We would like to discuss how the edges modes, whose number is expected to be measured by the area formula $S=\frac{A}{4G_N}$, emerge from string theory as a theory of quantum gravity. For this purpose, we will focus on the tree level (or genus zero) closed string contributions, which was pioneered by the paper \cite{Susskind:1994sm}. Refer to \cite{Dabholkar:1994ai,He:2014gva,Witten:2018xfj} for closed string results at one-loop level and to \cite{Balasubramanian:2018axm}  for results for open strings. We also simply assume that the spacetime is flat i.e. $R^{D}$, where $D=26$ in bosonic string and $D=10$ in superstring. We denote the spacetime coordinate by $(x_0,x_1,\ddd,x_{D-1})$. 
Then we consider the entanglement entropy between half spaces $A$ and $B$ at the time $x_0=0$, where $A$ and $B$ are simply defined by
$x_1>0$ and $x_1<0$, respectively. In this case, the replica method calculation of entanglement entropy is equivalent to the 
calculation of entropy in the Rindler space.

First we would like to note that it is well-known that the low energy effective action of closed string theory  
includes the Einstein-Hilbert action. For example, in the sigma model approach to closed string theory 
\cite{Fradkin:1985ys,Tseytlin:1988tv,Tseytlin:2000mt}, the tree level effective action of string theory $I_{st}$ is related to the genus zero (i.e. sphere) partition function $Z(S^2)$ via 
the following formula \cite{Tseytlin:2000mt}
\ba
I_{st}=-\frac{d Z(S^2)}{d\log \ep}\Biggr |_{\ep=1},   \label{relationst}
\ea
where $\ep$ is the UV cut off in the world-sheet theory on the sphere.  
The radius of the sphere is taken to be unit.  Therefore, we can perfectly derive the area formula of entropy 
in flat spacetime via the Euclidean approaches such as  the deficit angle method
and the one in earlier sections of the present paper using the Hayward term. 
Note that stringy corrections (or $\al$ corrections) are vanishing in flat spaces at genus zero. 

\begin{figure}
  \centering
  \includegraphics[width=6cm]{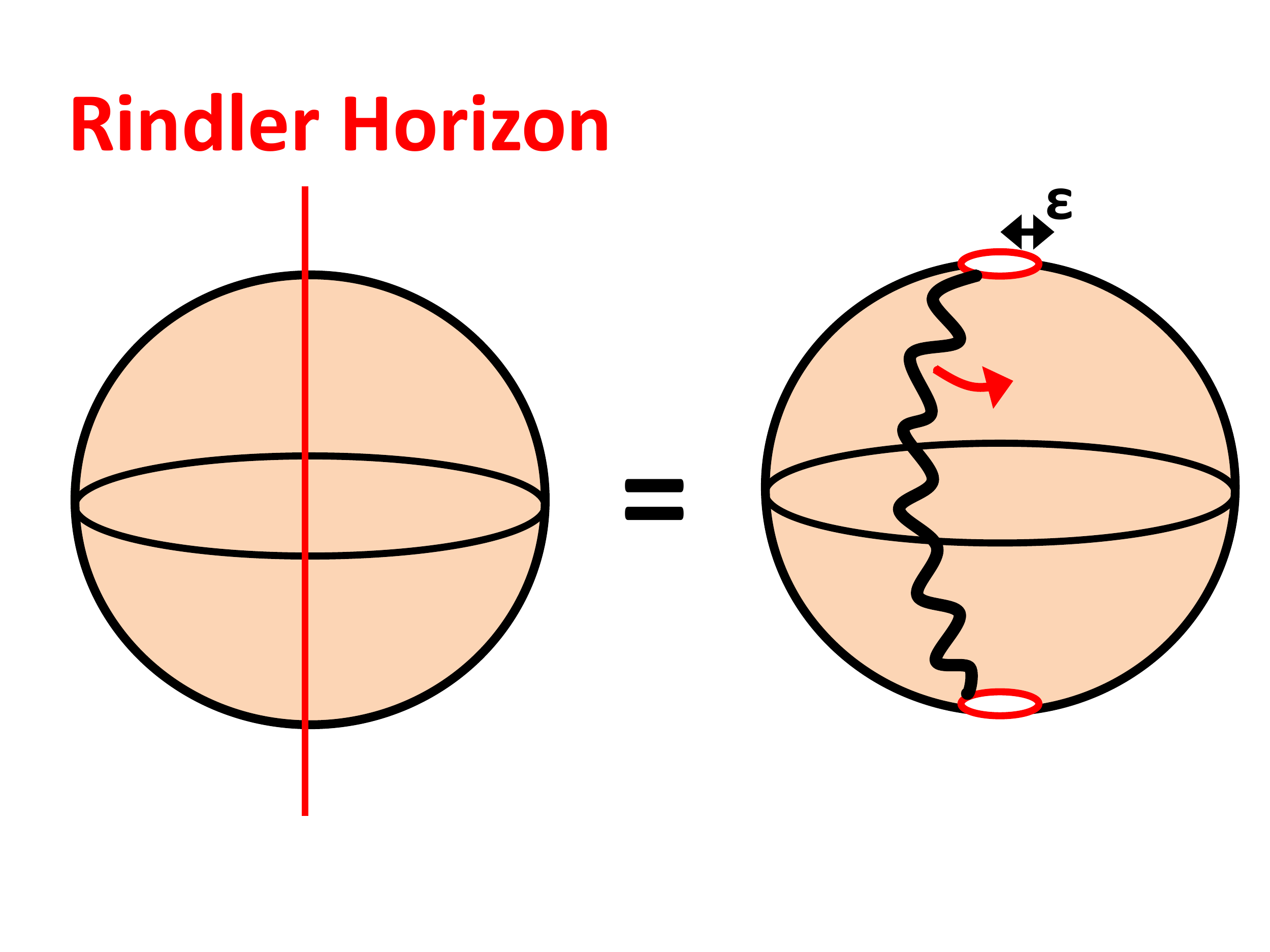}
  \caption{The sphere partition function of closed string world-sheet with north and south pole pinned at the Rindler horizon (left) can be equivalently described by the one-loop partition function of open string which connects the two poles (right). }
\label{openf}
\end{figure}

However, this argument does not answer the fundamental question where the gravity edges modes come from in string theory as a theory of quantum gravity. An important observation made in \cite{Susskind:1994sm} is that the edge modes may come from special world-sheet configurations which wind around the Rindler horizon $x_0=x_1=0$. In the replica computation of $\mbox{Tr}\rho_A^n$, the spherical world-sheet wraps $n$ times around the Rindler horizon such that the north and south pole of the sphere are fixed on the horizon. We can view this world-sheet as a one-loop diagram of an open-string which connects the north and south pole as depicted in Fig.\ref{openf}.

One might think that instead of considering replicated spacetimes, we may calculate the entropy from genus zero partition functions on $Z_N$ orbifolds (or the Melvin twist geometry \cite{Russo:1995tj,Takayanagi:2001jj}), as employed in the case of genus one partition function \cite{Dabholkar:1994ai,He:2014gva,Witten:2018xfj}, via the formula 
\ba
S'=-\frac{\de}{\de N}\left[I_{st}(C/Z_N\times R^{D-2})-\frac{1}{N}I_{st}(R^D)\right]\Biggr|_{N=1},  \label{xsxx}
\ea
where $I_{st}$ is calculated from the sphere partition function as (\ref{relationst}). In this correspondence, we regard the replica number $n$ as $1/N$. However, we immediately find that this quantity $S'$ in the orbifold CFT is simply vanishing. This is because the orbifold projection affects only partition functions 
with genus one or higher. For genus zero, we simply have $I_{st}(C/Z_N)-\frac{1}{N}I_{st}(R^2)=0$, because the orbifold projection acts in a rather trivial way for the genus zero partition function as opposed to the situation in the genus one case. This is consistent with the argument of \cite{Dabholkar:2001if}, where it was pointed out that the Einstein equation is satisfied at the fixed point of $C/Z_N$ owing to the localized tachyon potential, which means the on-shell condition i.e. 
conformal invariance on the worldsheet $\frac{d Z(S^2)}{d\log \ep}=0$.

On the other hand, if we try to employ the replica method with integer $n$, we need to calculate the partition function of the sigma model whose target space is the replicated geometry, which does not seem to be tractable at present. 

Instead of performing these closed string computations,  below we would like to evaluate the sphere partition function from the open string viewpoint. Indeed, we will see that our result implies that such configurations describe the gravitational edge modes.

\subsection{Edge Mode Partition Function in Open Bosonic String}

We would like to evaluate the sphere partition function from the open string viewpoint.
In this subsection we focus on bosonic string theory. The world-sheet theory is described by 
$26$ scalars $X^0,X^1,\ddd,X^{25}$ and $b,c$ ghost.
We write the metric of the unit sphere $S^2$ as
\ba
S^2:\ \ ds^2=d\theta^2+\sin^2\theta d\vp^2,\ \ (0\leq \vp\leq 2\pi,\ \ 0\leq \theta\leq \pi).  \label{spqw}
\ea
We assume that the north pole $\theta=0$ and the south pole $\theta=\pi$ are situated at the Rindler horizon $x_0=x_1=0$. We would like to view the sphere partition function as the one-loop cylinder partition function of open string.  At the open string end points, we choose the Dirichlet boundary condition for $X^0,X^1$ and the Neumann boundary condition for $X^2,\ddd,X^{25}$. Notice that we choose the Neumann boundary condition (rather than Dirichlet one) for the latter because we do not need to impose the continuity at the horizon for edge modes. 

\begin{figure}
  \centering
  \includegraphics[width=6cm]{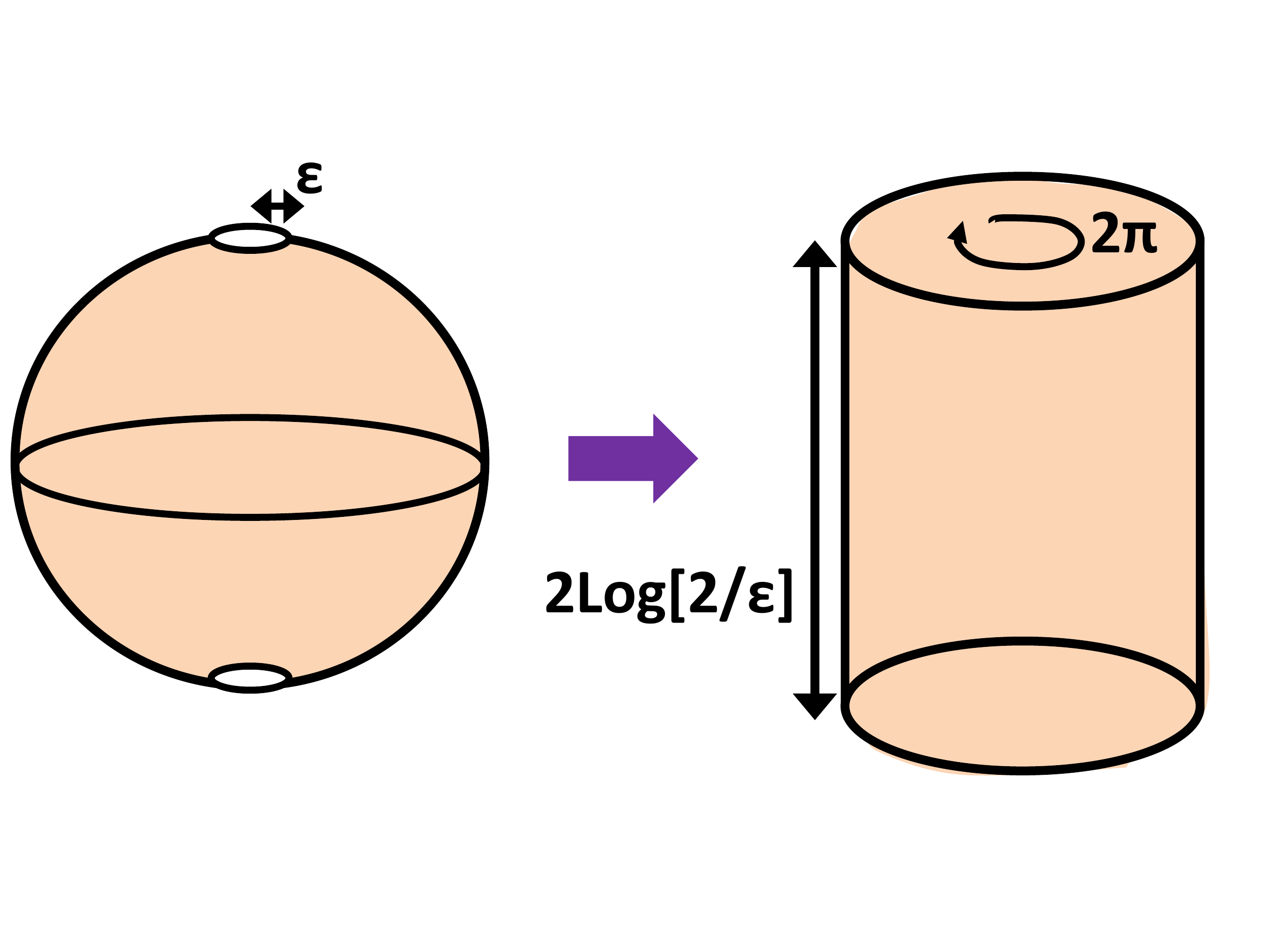}
  \caption{Conformal mapping from a unit radius sphere with two holes into a cylinder.}
\label{cmaps}
\end{figure}

For this, we put an infinitesimally small hole (radius $\ep$) on the north and south pole of the unit sphere $S^2$ as in Fig.\ref{cmaps}. This is described by the metric (\ref{spqw}) with the restriction 
$\ep\leq \theta\leq \pi-\ep$.   We map this sphere into a plane with the complex coordinate $w$ (using the usual stereographic map) such that
\ba
w=\frac{2\cos\frac{\theta}{2}}{\sin\frac{\theta}{2}}\cdot e^{i\vp}.
\ea
Finally the $w-$plane is mapped into a cylinder via the coordinate transformation $w=e^{\tau+i\sigma}$, where the cylinder coordinate $(\tau,\sigma)$ takes the following values
\ba
0\leq \sigma\leq 2\pi,\ \ \ \log\ep\leq \tau\leq \log(4/\ep),
\ea
as depicted in Fig.\ref{cmaps}.

Note that the metric on the sphere is related to the cylinder metric by the Weyl transformation 
\ba
&& ds^2=d\theta^2+\sin^2\theta d\phi^2=e^{2\phi}(d\tau^2+d\sigma^2), \no
&& e^{\phi}\equiv \frac{|w|}{1+|w|^2/4}.  \label{lifg}
\ea

Before we analyze bosonic string theory, consider a unitary CFT with the central charge $c$ as 
a simpler exercise. In this case, following the general result  (\ref{spherepe}), 
 we know that the sphere partition function should behave as 
\ba
Z(S^2)\propto e^{-\frac{c}{3}\log\ep},  \label{sphereP}
\ea
where the $\log\ep$ dependence is fixed by conformal anomaly. This is reproduced by the cylinder partition function $Z(Cyl)$ as follows:
\ba
Z(S^2)=e^{S_L[\phi]}\cdot Z(Cyl),  \label{deomps}
\ea
where $S_L[\phi]$ is the Liouville action \cite{Polyakov:1981rd} given by
\be
S_L[\phi]=\frac{c}{24\pi}\int d\tau d\sigma[(\de_\tau\phi)^2+(\de_\sigma \phi)^2+e^{2\phi}].
\ee
This Liouville term arises via the conformal anomaly for the Weyl rescaling (\ref{lifg}) and 
is estimated as follows:
\ba
S_L[\phi]=\frac{c}{6}\log\frac{2}{\ep}.
\ea
On the other hand, the cylinder amplitude behaves as 
\ba
Z(Cyl)=\mbox{Tr}[e^{-2\pi t H_{open}}]=\la 0|e^{-s(L_0+\bar{L}_0-\frac{c}{12})}|0\lb \simeq e^{\frac{c}{12}s}=e^{\frac{c}{6}\log\frac{2}{\ep}}, 
\ea
where we introduced the moduli parameter $t$ and $s$ such that
\ba
s=\frac{\pi}{t}=2\log\frac{2}{\ep}.
\ea
Also $H_{open}$ is the Hamiltonian in the open string channel and $(L_0,\bar{L}_0)$ describe the conformal dimensions in the closed string channel as usual. It is easy to confirm that (\ref{sphereP}) is reproduced from (\ref{deomps}).

Now we move back to the open bosonic string. This corresponds to coupling the $b,c$ ghost CFT to the 
$c_X=26$ unitary CFT. To calculate the sphere partition function, we need to insert three left-moving
and three right-moving $c$-ghost operators as $\la c\ti{c}c\ti{c}c\ti{c} \lb$. Importantly, for a standard normalization which leads to a finite sphere partition function,
we need to insert them in generic middle points to avoid the singular factor due to the Weyl rescaling 
(\ref{lifg}). On the other hand, the standard cylinder amplitude of open string \cite{Pol} corresponds to the insertions of a pair $c\ti{c}$ into each boundary of the cylinder. Therefore if we naively regard $Z(Cyl)$
as the standard open string amplitude, we will obtain the extra divergent factor due to the Weyl rescaling\footnote{To see this remember that the ghost $c$ has the chiral dimension $-1$.}:
\be
e^{-2\phi_N}\cdot e^{-2\phi_S}\simeq \frac{1}{\ep^4}\propto e^{2s},  \label{elimt}
\ee
where $\phi_N$ and $\phi_S$ are the value of $\phi$ at the north boundary $\theta=\ep$ and 
the south boundary $\theta~=\pi-\ep$ of the cylinder. Also note that the Liouville action contribution (\ref{deomps}) is trivial i.e. $S_L[\phi]=0$ because the total central charge is vanishing $c_X+c_{bc}=0$.

 In this way, eventually we obtain the sphere partition function with two infinitesimally small hole:
\ba
Z(S^2)_{edge}&=&
g_s^{-2}\lim_{\ep\to 0} \left[e^{-2s}\mbox{Tr}[e^{-2\pi H_{open}t}]\right],\no
&=&A_{24}g_s^{-2} \lim_{\ep\to 0} \left[ \left(8\pi^2 \al t \right)^{-12}\cdot  e^{-2s}\cdot \eta(it)^{-24}\right],\no
&=&A_{24}g_s^{-2} \lim_{\ep\to 0} \left[ \left(8\pi^2 \al \right)^{-12}\cdot  e^{-2s}\cdot \eta(is/\pi)^{-24}\right],\no
&=& A_{24}g_s^{-2}\cdot \left(8\pi^2 \al \right)^{-12},  \label{stringpar}
\ea
where $g_s$ is the string coupling constant, which arises as we consider the genus zero partition function,\footnote{
If we purely consider the open string picture, we might not have $g_s$ as it is a cylinder amplitude. In that case we may 
obtain $1/g_s^2$ factor from the Chan-Paton degrees of freedom as suggested in \cite{Lin:2017uzr}. However, 
in this paper our strategy is to reinterpret the closed string computations in terms of open strings.}
  and $A_{24}$ is the total area of the Rindler horizon. 
Also we employed the modular transformation 
\ba
\eta(i/t)=\s{t}\ \eta(it),
\ea
of the eta function $\eta(it)=e^{-\frac{\pi}{12}t}\prod_{m=1}^\infty (1-e^{-2\pi mt})$. Remember also that we multiplied $e^{-2s}$ to remove the unwanted divergent factor (\ref{elimt}). Note that we do not know the precise normalization of $Z(S^2)$ suitable for our problem as the location of $c$ ghosts can move in many ways with the current knowledge. Therefore we ignore this $O(1)$ undetermined normalization factor in (\ref{stringpar}).

In summary, we find 
\ba
Z(S^2)_{edge}=n_{26}\cdot \frac{A_{24}}{G_N},
\ea
where $G_N\propto g_s^2\al^{12}$ is the Newton constant in bosonic string and $n_{26}$ is an $O(1)$ 
undetermined factor.

\subsection{Edge Mode Partition Function in Open Superstring}

Let us turn to superstring. Consider an evaluation of genus zero partition function from open superstring viewpoint. In the same way as in the bosonic string case, we impose the Dirichlet boundary condition in $X^0,X^1$ and the Neumann one in $X^2,\ddd,X^9$, similarly for their super-partners. The insertion of ghost operators now look like $\la c\ti{c}\delta(\gamma)\delta(\ti{\gamma})\cdot c\ti{c}\cdot  \la c\ti{c}\delta(\gamma)\delta(\ti{\gamma})\lb$, where $\delta(\gamma)$ describes a vertex operator in the $(-1)$ picture.  Since the conformal dimension of $c\delta(\gamma)$ is $-\frac{1}{2}$, we need to multiply 
$e^{-s}$ (instead of $e^{-2s}$ for the bosonic string) to remove unwanted divergence.  Since the small disk 
path-integral corresponds to the vacuum state in the matter CFT via the state/operator mapping,  our cylinder amplitude corresponds to the propagation of the tachyon state in the NSNS sector of the closed string. Note that we do not impose the GSO projection as we are looking at the genus zero partition function. 
The IR divergence of the closed string tachyon is canceled by the multiplication of the factor $e^{-s}$.
In the open string channel we only have NS sector without GSO projection. Though this might be analogous to the non-BPS D-brane \cite{Bergman:1998xv,Sen:1998tt}, they are different in that the latter has both NS and R sector, while the former does only NS sector. 

In this way, the edge mode contributions to the sphere partition function in superstring is evaluated as follows:  
\ba
Z(S^2)_{edge}&=&
g_s^{-2}\lim_{\ep\to 0} \left[e^{-s}\mbox{Tr}[e^{-2\pi H_{open}t}]\right],\no
&=& A_{8}g_s^{-2}\cdot \left(8\pi^2 \al \right)^{-4},
\ea
where we omitted the details of modular transformation as they are parallel with the bosonic string.

Therefore we obtain
\ba
Z(S^2)_{edge}=n_{10}\cdot \frac{A_{8}}{G_N},
\ea
where $n_{10}$ is again an undertermined $O(1)$ constant.

\subsection{Possible Interpretation as Gravitational Entropy}

In terms of edge modes, which propagate around the Rindler horizon,
we would like to interpret the sphere partition function  $Z(S^2)_{edge}$
as that of the edge modes
\ba
Z(S^2)_{edge}=\mbox{Tr}_{edge}[e^{-2\pi H_{edge}t}].,  \label{edgest}
\ea
where we introduced $H_{edge}=H_{open}+\frac{1}{\pi t}\log g_s +\frac{s}{2\pi t}$.
Since we are taking the limit $\ep\to 0$ i.e. $t\to 0$, (\ref{edgest}) shows that the number of 
edge modes is $n_{10} \frac{A_8}{G_N}$, where $A_8$ is the area of Rindler horizon. 
However, notice that this world-sheet partition function 
is at the first quantized level as usual in string theory. 
The second quantized partition function is given by the exponential $e^{Z(S^2)_{edge}}$. 
This leads to the estimation of the number of second quantized edge modes
$e^{n_{10}\frac{A_8}{G_N}}$. This fits nicely with the expected answer $e^{\frac{A_8}{4G_N}}$ 
based on the gravitational entropy up to the numerical factor $n_{10}$. This strongly suggests that 
our identification of open strings with edge modes is correct.

\section{Conclusions}

In this work, we have pointed out the role of Hayward term, which is corner contributions in gravity action, as gravity edge modes. Interestingly, the Hayward term explains the area term in the gravitational entropy for any spacetimes, including asymptotically flat ones. Based on the Hayward term, we explained the pasting rule of gravity action which gives rise to the non-additivity of the action in the presence of wedges. 
This non-additivity due to the Hayward term directly suggests that we need to insert the so-called area operator in order to account for the factorized Hilbert space. Recently, the necessity of such operator insertion is discussed explicitly in the context of two-dimensional gravity\cite{Lin:2018xkj,Jafferis:2019wkd}. The Hayward term would reflect a such necessity in more generic sense. 

We have also presented two different interpretations of setups where gravity systems are divided into subsystems. One is based on a generalization of 
holographic and the other is based on the AdS/BCFT. In the former, we argued that the Hayward term contribution describes the partition function on the extra 
boundary. We also showed that the wave functional viewpoint of gravitational theory and conjectured 
gravity entanglement formula reproduce the Hayward term correctly. 
In the latter, we found that the gravitation entropy associated with the area of the separation surface counts the number of possible boundary conditions 
for the CFT with the boundary. Moreover, we showed that the Hayward term represents the conformal anomaly in another setup of AdS/BCFT. 
All these show that the gravitational edge modes show up universally as dynamical degrees of freedom on edges of subsystems in holography. 

We also noted that these structures of gravitational subsystems turn out to be quite analogous to the ones in gauge theories. We can understand the origins of the area term  
because the entanglement entropy in gauge theories depends on the choice of the gauge fixing across the boundaries, which is also the case for gravitational entropy. It is natural because, for the observers in the subsystem, such gauge degrees of freedom on the boundary become physical as like the asymptotic symmetry. However, it is still not so clear what is the proper choice of the boundary condition to reproduce the area term. 

Toward the above question, we studied a possible interpretation of gravity edge modes as the degrees of freedom in the string theory. We evaluated the sphere partition function of closed string with punctures (equivalent to one-loop cylinder partition function of open string) and saw a qualitative agreement with the finite area term as the open string degrees of freedom. 
Our argument strongly suggests that the stringy degrees of freedom would account for the gravity edge modes. It means that the ``gauge-variant'' subspace does indeed make sense from the string theory viewpoint. Further clarification of this idea is a challenging but very interesting future direction.

\section*{Acknowledgements}
We thank Ibrahim Akal, Pawel Caputa, Teppei Shimaji and Zixia Wei for useful discussions. 
KT and TT are supported by the Simons Foundation through the ``It from Qubit'' collaboration.
KT is supported by the JSPS Grant-in-Aid for Research Activity start-up 19K23441.
TT is supported by World Premier International Research Center Initiative (WPI Initiative) 
from the Japan Ministry of Education, Culture, Sports, Science and Technology (MEXT). 
TT is also supported by JSPS Grant-in-Aid for Scientific Research (A) 16H02182 and 
by JSPS Grant-in-Aid for Challenging Research (Exploratory) 18K18766.



\end{document}